\documentclass[11pt, a4paper]{article}

\pdfoutput=1

\usepackage{jcappub}
\usepackage{xfrac}
\usepackage{times}
\usepackage{etoolbox}
\usepackage{bm}
\renewcommand{\vec}[1]{\mathbf{#1}}
\usepackage[utf8]{inputenc}

\makeatletter
\patchcmd{\maketitle}{\@fpheader}{}{}{}
\makeatother
 
\title{Relating quantum mechanics and kinetics of neutrino oscillations}

\author{A. Kartavtsev\\}

\affiliation{P. G. Demidov Yaroslavl State University, Sovietskaya 14, 150003 Yaroslavl, Russia}
 
\abstract{Simultaneous treatment of neutrino oscillations and collisions in astrophysical environments requires the use 
of (quantum) kinetic equations. Despite major advances in the field of quantum kinetics, the structure of the kinetic equations 
and their consistency with the uncertainty principle are still debated. The goals of the present work are threefold. First, it 
clarifies the structure of the Liouville term in the presence of mixing. Second, we derive evolution equation for neutrinos 
propagating in vacuum or matter from the Schr\"odinger equation and show that in the relativistic limit its form 
matches the form of the (collisionless part of the) kinetic equation derived by Sigl and Raffelt. 
Third, by constructing solutions of the evolution equation from the known solutions of the Schr\"odinger equation, we show
that the former also admits solutions consistent with the uncertainty principle and accounts for neutrino wave packet 
separation. The obtained results speak in favor of a (quantum) kinetic approach to the analysis of neutrino propagation 
in exploding supernovae where neutrino oscillations and collisions, as well as the effect of wave packet separation, might 
be equally important.}


\begin{document}

\maketitle
\flushbottom
\newpage

\section{\label{sec:introduction}Introduction}

The disappearance of solar neutrinos reported by Ray Davis Jr and his collaborators \citep{Davis:1968cp} was the first 
hint of the existence of flavor neutrino oscillations. Shortly thereafter an expression for the survival probability of $\nu_e$ was 
obtained by Gribov and Pontecorvo \cite{Gribov:1968kq}. While propagation of the low-energy fraction of solar neutrinos 
detected by the Gallium experiments \cite{Gavrin:2007wc} is almost unaffected by the matter effects \cite{Strumia:2006db}, 
forward scattering off the background matter strongly affects propagation of the high-energy fraction  observed by the water
Cherenkov experiments \cite{Hosaka:2005um}. However, propagation of even the high energy neutrinos in the sun's interior 
is almost collisionless and can be described by the Schr\"odin\-ger equation for the neutrino wave function $\psi_i(t,\vec{x})$ 
or by the equivalent Liouville--von Ne\-u\-mann equation for the respective density matrix,
\begin{align}
\label{eq:densitymatrix}
\rho_{ij}(t,\vec{x})=\psi_i(t,\vec{x})\psi^{*}_j(t,\vec{x})\,,
\end{align}
see reference \cite{Akhmedov:2009rb} for a comprehensive review. As has been emphasized in references 
\cite{Hansen:2016klk,Akhmedov:2017mcc}, this appro\-ach is consistent with the uncertainty principle and convenient for 
analysis of wave packet separation.
 
On the other hand, neutrino collisions with particles of the ambient medium can play a dominant role in certain phases 
of supernovae evolution \cite{Sigl:1992fn}. Because particle number is conserved in quantum mechanics, description of  
neutrino production and absorption processes typically relies on the Boltz\-mann kinetic equation for the neutrino 
occupation numbers. Their quantum-mechanical counterpart is the Wigner function:
\begin{align}
\label{eq:wignerfunction}
\varrho_{ij}(t,\vec{x},\vec{p}) = \int d^3\vec{y} e^{-i\vec{p}\vec{y}} 
\psi_i(t,\vec{x}+\vec{y}/2)\psi^{*}_j(t,\vec{x}-\vec{y}/2)\,.
\end{align}
The Boltzmann equation in its classical form is not capable of describing neutrino oscillations. Thus,
neither the Liouville--von Neumann nor the Boltzmann equation can consis\-tently describe neutrino propagation from the 
neutrinosphere, where collisions dominate, to the outer layers, dominated by oscillations. The difficulty of ``marrying'' 
oscillations and collisions is evident already from the fact that the Liouville--von Neumann and Boltzmann equations operate 
with different objects, the density matrix and occupation numbers respectively. One of the first attempts to incorporate oscillations into the kinetic equations was made by Dolgov \cite{Dolgov:1980cq} who replaced, for each momentum mode, the 
neutrino occupation numbers by matrices in flavor space. Whereas diagonal entries of these matrices represent the usual occupation 
numbers, the off-diagonals correspond to coherence between the neutrino flavor eigenstates. In reference \cite{Sigl:1992fn} 
Sigl and Raffelt derived kinetic equation for the matrices of occupation numbers featuring correct scattering, oscillation, and, 
last but not least, Liouville term, applicable in the physically relevant relativistic limit. However, their result is sometimes viewed 
with a degree of skepticism, mainly because the derived kinetic equation is believed to be in conflict with the uncerta\-inty 
principle as well as unable to account for the effect of wave packet separation. 

A technically different approach to the analysis of neutrino oscillations was employed by 
Yama\-da in reference \cite{Yamada:2000za} and by Vlasenko, Fuller, and Cirigliano in references 
\cite{Vlasenko:2013fja,Cirigliano:2014aoa,Vlasenko:2014bva,Blaschke:2016xxt,Richers:2019grc}. Instead of 
analyzing the density matrix or the matrix of densities they used Kadanoff-Baym equations for the neutrino 
Wightman propagator. Its quantum-mechanical counterpart is the two-point correlator:
\begin{align}
\label{eq:correlator}
\varrho_{ij}(t,\vec{x},\epsilon,\vec{p}) =\int d\tau e^{i\epsilon\tau}\int d^3\vec{y} e^{-i\vec{p}\vec{y}} 
\psi_i(t+\tau/2,\vec{x}+\vec{y}/2)\psi^{*}_j(t-\tau/2,\vec{x}-\vec{y}/2)\,.
\end{align}
The obtained quantum kinetic equation is capable of describing neutrino flavor and spin oscillations, as well as neutrino 
collisions with the ambient medium. However, their results have not yet received the deserved attention in the community, 
mainly because of the high technical complexity of the used formalism. 

The goal of the present work is to establish connection between these three approaches to neutri\-no oscillations and present 
arguments in favor of the (quantum) kinetic approach to the analysis of neutrino propagation in supernovae, where 
oscillations and collisions, as well as the effect of wave packet separation, might be equally important. To this end we first 
consider neutrino oscillations in vacuum in section \ref{sec:liouvilleterm}. In subsection \ref{sec:quantummechanics} we briefly 
review the quantum-mechanical approach. 
In subsection \ref{sec:quantumkineticsvac} we derive evolution equation for the 
neutrino two-point correlator from the Schr\"odinger equation. Its form matches the form of the quantum kinetic 
equation in vacuum. We also construct its solutions from the known
solutions of the Schr\"odinger equation in vacuum and analyze the shell structure of the resulting two-point correlator. In addition 
to the mass shells $\epsilon = \omega_i$ we identify middle shells $\epsilon = (\omega_i+\omega_j)/2$, that are responsible for
neutrino oscillations. In subsection \ref{sec:kinetics} we derive evolution equation for the neutrino Wigner function 
from that for the two-point correlator in the on-shell approximation (that necessarily includes the middle shells)
and clarify the structure of its Liouville term in the presence of mixing. For relativistic neutrinos 
its form matches the (vacuum limit of the) kinetic equation of Sigl and Raffelt. Considering Gaussian and Lorentzian initial conditions
we  show that it admits solutions that are consistent with the uncertainty principle and account for the wave packet separation.
In section \ref{sec:quantumkinetics} these results are generalized to collisionless neutrino propagation 
in matter. In subsection \ref{sec:quantumkineticsmed} we derive, to the first order in the gradient expansion,
evolution equation for the neutrino two-point correlator valid also for non-relativistic neutrinos. Studying it term by term we single out contributions relevant in the physically interesting relativistic limit. In subsection \ref{sec:kineticsmed} we derive 
evolution equation for the Wigner function from the relativistic limit of the evolution equation for the 
two-point correlator. Its form matches the (collisionless limit of the) kinetic equation of Sigl and Raffelt. Constructing the Wigner 
function from the formal solution of the Schr\"odinger equation we show that it admits solutions that are consistent with the 
uncertainty principle and account for neutrino wave packet separation also for neutrino propagation in matter. In section 
\ref{sec:conclusions} we summarize our results. Finally, appendices \ref{sec:covliouville} to \ref{sec:quantumkineticeq} 
contain supplementary 	technical material.

\section{\label{sec:liouvilleterm}Neutrino propagation in vacuum}

In this section we consider neutrino propagation in vacuum and establish a connection between three technically different 
approaches to description of flavor neutrino oscillations. In subsection \ref{sec:quantummechanics} we briefly review the  
quantum-mechanical approach operating with the neutrino wave function $\psi_i(t,\vec{x})$ or density matrix $\rho_{ij}(t,\vec{x})$. 
In subsection \ref{sec:quantumkineticsvac} we derive evolution equation for the two-point correlator 
$\varrho_{ij}(t,\vec{x},\epsilon,\vec{p})$ from the Schr\"odinger equation. Its form matches (the vacuum limit
of) the quantum kinetic equation. In subsection \ref{sec:kinetics}  we derive evolution equation for 
the Wigner function $\varrho_{ij}(t,\vec{x},\vec{p})$ from the evolution for the two-point correlator using on-shell 
approximation. In the physically relevant relativistic limit its form matches (the vacuum limit of) the kinetic equation of Sigl 
and Raffelt.

\subsection{\label{sec:quantummechanics}Description in terms of the density matrix}
We first  briefly review the standard quantum mechanical wave packet approach to  neutrino oscillati\-ons. A comprehensive 
review and analysis of  subtleties related to the neutrino production, propagation, and detection can be found in reference
\cite{Akhmedov:2009rb}.

\paragraph{Schr\"odinger equation for neutrino wave function.} 
Neutrinos produced in charged-current weak interaction processes are considered to be in a flavor eigenstate 
$\nu_\alpha \in (\nu_e,\nu_\mu,\nu_\tau)$ \cite{Akhmedov:2010ms}.  Up to small corrections \cite{Cozzella:2018zwm}
the latter is a linear superposition of the neutrino mass eigenstates $\nu_i\in (\nu_1,\nu_2,\nu_3)$. In terms of the 
wave functions $\psi_\alpha\! =\!	 \sum_i  U_{\alpha i}\,\psi_i$. Evolution of the mass eigenstates is governed 
by the Schr\"odinger equation,
\begin{align}
\label{eq:schroedinger}
i\partial_t \psi_i(t,\vec{x})={\sf H}_{ij}(t,\vec{x}) \psi_j(t,\vec{x})\,.
\end{align}
In vacuum ${\sf H}_{ij}(t,\vec{x})={\sf K}_{ij}(\vec{x})$ with the kinetic energy operator given by
${\sf K}_{ij}=\delta_{ij}(-\partial^2_{\vec{x}}+m_i^2)^\frac12$ (note that throughout this paper we work in the mass
basis). As can be verified by substitution, resulting solution of the Schr\"odinger equation is given by \cite{Hansen:2016klk}
\begin{align}
\label{eq:wavepacket}
\psi_i(t,\vec{x})=\int\frac{d^3\vec{p}}{(2\pi)^3} \psi_i(0,\vec{p})e^{-i(\omega_i(\vec{p})t-\vec{p}\vec{x})}\,,
\end{align}
with  $\omega_i(\vec{p})\equiv(\vec{p}^2+m_i^2)^\frac12$. 
  
The initial conditions are encoded in $\psi_i(0,\vec{p})$. For the reason of computational simplicity it is common to assume 
that the mass eigenstates constituting the produced flavor state are described in the momentum space by Gaussian  wave 
packets \cite{Hansen:2016klk,Akhmedov:2017mcc}, 
\begin{align}
\label{eq:gaussianwp}
\psi_i(0,\vec{p}) \propto \exp\left(-\frac{(\vec{p}-\vec{p}_w)^2}{4\sigma_p^2}\right)\,,
\end{align} 
where $\sigma_p$ is  the momentum uncertainty of the produced neutrino state and $\vec{p}_w$ -- its characteristic 
momentum. Following reference \cite{Hansen:2016klk} we neglect the difference in the shape of $\psi_i(0,\vec{p})$ 
for different mass eigenstates.  As has been shown in reference \citep{Akhmedov:2010ms} by analyzing neutrino 
oscillations in the QFT framework (see also references \cite{Grimus:2019hlq,Grimus:2003es,Grimus:1998uh,Grimus:1996av}), expression \eqref{eq:gaussianwp} is oversimplified and accounts neither for the
difference of the momentum uncertainties in different directions, nor for the contribution of the energy uncertainty. 
However,  it is  sufficient for the purposes of this work, where it is used solely for illustration. 
For small $\sigma_p$ the corresponding coordinate-space wave function is well approximated by 
\begin{align}
\label{eq:psigaussian}
\psi_i(t,\vec{x})\propto \exp\left(-\frac{(\vec{v}_i(\vec{p}_w)t-\vec{x})^2}{4\sigma_x^2}\right) 
e^{-i(\omega_i(\vec{p}_w)t-\vec{p}_w\vec{x})}\,,
\end{align}
with  $\vec{v}_i(\vec{p})\equiv \partial_\vec{p}\omega_i(\vec{p})$ being velocity of the respective mass eigenstate.

Equation \eqref{eq:gaussianwp} states	that the momentum-space wave function is localized in the vicinity of $\vec{p}_w$
within the $\sim \sigma_p$ range. Its coordinate-space counterpart is localized in the vicinity of $\vec{x}=\vec{v}t$ within 
the $\sim \sigma_x$ range, where the latter is defined by the uncertainty relation, $\sigma_x\sigma_p=\frac12$. That is, in 
agre\-e\-ment with the uncertainty principle, delocalization in the momentum space results in localization in the coordinate 
space, and vice versa. 

\paragraph{Liouville--von Neumann equation for neutrino density matrix.}
Since one is typically interested in the probability of detecting a neutrino of a particular flavor $\alpha$, 
$P_\alpha=|\psi_\alpha|^2=\sum_{ij}U_{\alpha i} U^*_{\alpha j}\,\psi_i\psi^*_j=\sum_{ij}U_{\alpha i} U^*_{\alpha j}\,\rho_{ij}$	,
a convenient quantity for tracking the neutrino evolution is the 	density matrix  defined in equation  \eqref{eq:densitymatrix}. 
Using the latter we find that the Schr\"odinger equation translates, without any approximations, into a closed-form 
Liouville--von Neumann equation for the density matrix,
\begin{align}
\label{eq:LiovilleVonNeumann}
\partial_t \rho(t,\vec{x}) = -i[{\sf H}(t,\vec{x}),\rho(t,\vec{x})]\,.
\end{align}
Note that as in reference \cite{Stirner:2018ojk} $\rho$ and $\sf H$ without generation indices denote matrices in the	 
mass basis.
Using the approximate solution equation \eqref{eq:psigaussian} we obtain for the density matrix  
\cite{Akhmedov:2017mcc,Hansen:2016klk}
\begin{align}
\label{eq:densitymatrixgaussian}
\rho_{ij}(t,\vec{x}) \propto  \exp\biggl(-\frac{(\bar{\vec{v}}_{ij}(\vec{p}_w)t-\vec{x})^2}{2\sigma_x^2}\biggr) 
\exp\biggl(-\frac{\Delta\vec{v}^2_{ij}(\vec{p}_w)t^2}{8\sigma^2_x}\biggr)e^{-i(\omega_i(\vec{p}_w)-\omega_j(\vec{p}_w))t}\,,
\end{align}
where $\bar{\vec{v}}_{ij}=\frac12(\vec{v}_i+\vec{v}_j)$ and $\Delta\vec{v}_{ij}\equiv \vec{v}_i-\vec{v}_j$. 
Note that because equation \eqref{eq:psigaussian} only approximately solves the Schr\"odinger equation, 
equation \eqref{eq:densitymatrixgaussian} only approximately solves the Liouville--von Neumann equation.
Equation \eqref{eq:densitymatrixgaussian} states that components of the density matrix are localized in the vicinity of 
$\vec{x}=\vec{\bar{v}}_{ij}t$ within the $\sim\sigma_x$ range, thereby reflecting the uncertainty principle. In addition, 
it implies that the off-diagonal elements of the density matrix experience a suppression proportional to the difference of 
the respective group velocities. This is a manifestation of kinematical decoherence \cite{Akhmedov:2009rb,Stirner:2018ojk} 
caused by wave packet separation: because wave packets of the neutrino mass eigenstates have finite spatial size 
$\sim \sigma_x$ and propagate with different group velocities they separate in the course of neutrino propagation. 

\subsection{\label{sec:quantumkineticsvac}Description in terms of the two-point correlator}
The definition of the density matrix $\rho_{ij}(t,\vec{x})$ requires the two underlying wave functions to be computed at the same point 
of the space-time, thereby reducing the number of `degrees of freedom' from four (two time and two space arguments) to 
two (one time and one space argument). This reduction is avoided in the two-point correlator $\varrho_{ij}(t,\vec{x},\epsilon,\vec{p})$
that trades the four arguments of 	the wave functions for another four -- the central time, central coordinate, energy, 
and momentum -- thereby providing a more granular description of the neutrino state than the density matrix.

As we demonstrate below, in vacuum the Schr\"odinger equation for the neutrino  wave function translates, without  any
approximations, into a closed-form evolution equation for the  two-point cor\-re\-lator. 
Its form matches (the vacuum limit of) the quantum kinetic equation.

The energy argument $\epsilon$ is independent of the momentum argument $\vec{p}$. This
means in particular, that the two-point correlator is not constrained to be on-shell. Of course, for typical initial conditions its 
diagonals do  peak in the vicinity of the respective mass shells, $\epsilon\sim \omega_i$. On the other hand,
the off-diagonals live on the intermediate shell(s) $\epsilon\sim (\omega_i+\omega_j)/2$. 

Given the arguments $\vec{p}$ and $\epsilon$ one can construct the usual 
relativistic velocity $\vec{v}=\vec{p}/\epsilon$  entering the Liouville term $\partial_t +\vec{v}\partial_\vec{x}$. The velocity 
does not carry any generation indices  because it is determined by the energy spectrum of the two-point correlator. Since the 
diagonals peak on the  mass shells, their propagation velocity matches that of the respective mass eigenstates. Because the 
off-diagonals live on the intermediate shell, their propagation velocity matches that of the center of momentum. As this is 
in no way obvious if one deals with the Wigner function or the density matrix, the question of the propagation velocity in the 
Liouville operator was debated in the literature recently \cite{Hansen:2016klk,Stirner:2018ojk}.

\paragraph{Evolution equation for neutrino two-point correlator.}
The two-point correlator corresponding to equation \eqref{eq:wavepacket} is derived by substituting the latter  into  equation
\eqref{eq:correlator} and integrating over $\tau$ and $\vec{y}$,
\begin{align}
\label{eq:correlatorfirststep}
\varrho_{ij}(t,\vec{x},\epsilon,\vec{p})  & = \int \frac{d^3\vec{k}}{(2\pi)^3}\,\frac{d^3\vec{q}}{(2\pi)^3} \,
(2\pi)^3\delta\biggl(\vec{p}-\frac{\vec{k}+\vec{q}}2\biggr) 
(2\pi)\delta\biggl(\epsilon-\frac{\omega_i(\vec{k})+\omega_j(\vec{q})}2\biggr) 
\nonumber\\
& \times \psi_i(0,\vec{k}) \psi^*_j(0,\vec{q})\,e^{-i(\omega_i(\vec{k})t-\vec{k}\vec{x})}e^{i(\omega_j(\vec{q})t-\vec{q}\vec{x})}\,.
\end{align}
Using the identity 
\begin{align*}
\omega_i(\vec{k})-\omega_j(\vec{q})=\frac{\omega^2_i(\vec{k})-\omega^2_j(\vec{q})}{\omega_i(\vec{k})+\omega_j(\vec{q})}
=(\vec{k}-\vec{q})\frac{\vec{k}+\vec{q}}{\omega_i(\vec{k})+\omega_j(\vec{q})}+
\frac{m_i^2-m_j^2}{\omega_i(\vec{k})+\omega_j(\vec{q})}\,,
\end{align*}
as well as relations $\vec{k}+\vec{q}=2\vec{p}$ and $\omega_i(\vec{k})+\omega_j(\vec{q})=2\epsilon$ implied by the
delta-functions, we can recast equation \eqref{eq:correlatorfirststep} as a product of a shape and a phase factor,
\begin{align}
\label{eq:shapephase}
\varrho_{ij}(t,\vec{x},\epsilon,\vec{p}) = g_{ij}(\vec{v}t-\vec{x},\epsilon,\vec{p})
\cdot\exp\biggl(-i\,\frac{m_i^2-m_j^2}{2\epsilon}\,t\biggr)\,.
\end{align}
The shape factor reads
\begin{align}
\label{eq:shapefactor}
g_{ij}(\vec{v}t-\vec{x},\epsilon,\vec{p})  & =   \int \frac{d^3\vec{k}}{(2\pi)^3}\,\frac{d^3\vec{q}}{(2\pi)^3} \,
(2\pi)^3\delta\biggl(\vec{p}-\frac{\vec{k}+\vec{q}}2\biggr) 
(2\pi)\,\delta\biggl(\epsilon-\frac{\omega_i(\vec{k})+\omega_j(\vec{q})}2\biggr) 
\nonumber\\
& \times \psi_i(0,\vec{k}) \psi^*_j(0,\vec{q}) \,e^{-i(\vec{k}-\vec{q})(\vec{v}t-\vec{x})}\,,
\end{align}
with the velocity $\vec{v}$ defined as $\vec{v}\equiv\vec{p}/\epsilon$. Because the shape factor is a function 
of  $\vec{v}t-\vec{x}$, it satisfies the Liouville equation,
\begin{align}
\label{eq:liouvilleshape}
(\partial_t+\vec{v}\partial_\vec{x})\,g_{ij}(\vec{v}t-\vec{x},\epsilon,\vec{p})=0\,.
\end{align}
Action of the Liouville operator on the two-point correlator, equation \eqref{eq:shapephase}, therefore results in
\begin{align}
\label{eq:liouvilleeq}
(\partial_t+\vec{v}\partial_\vec{x})\varrho_{ij}(t,\vec{x},\epsilon,\vec{p}) =
-i\frac{m_i^2-m_j^2}{2\epsilon}\varrho_{ij}(t,\vec{x},\epsilon,\vec{p})\,.
\end{align} 
Let us emphasize that the derivation of equations \eqref{eq:shapephase} and \eqref{eq:liouvilleeq} does not rely on any 
approximations, and that the form of equation \eqref{eq:liouvilleeq} is independent of the initial conditions. 

\paragraph{Propagation velocity.} Results similar to equation \eqref{eq:liouvilleeq} have
been obtained in a number of previous works \cite{Sirera:1998ia,Yamada:2000za,Cardall:2007zw,
Vlasenko:2013fja}. However, the authors quickly turned to the ultrarelativistic approximation setting $\epsilon=|\vec{p}|$ 
and leaving the meaning of $\epsilon$ unspecified. 

To clarify the role of $\epsilon$ we consider the plane wave limit, $\psi_i(0,\vec{p})\propto \delta(\vec{p}-\vec{p}_w)$. In this
limit the shape factor simplifies to 
\begin{align}
\label{eq:shapefactorplanewave}
g_{ij}(\vec{v}t-\vec{x},\epsilon,\vec{p}) \propto \delta(\vec{p}-\vec{p}_w)
\,\delta\biggl(\epsilon-\frac{\omega_i(\vec{p}_w)+\omega_j(\vec{p}_w)}2\biggr) \,,
\end{align}
where we have omitted the overall normalization  for brevity.  The second delta-function in equation \eqref{eq:shapefactorplanewave}
forces the diagonals  to the mass shell of the respective mass eigenstate, $\epsilon=\omega_i(\vec{p}_w)$. On the other hand, 
the off-diagonals live on the intermediate shell, $\epsilon=\sfrac{1\!}2(\omega_i(\vec{p}_w)+\omega_j(\vec{p}_w))$, see 
references \cite{Garny:2011hg,Kartavtsev:2015vto} for a related analysis in the context of leptogenesis. 

The non-trivial shell structure of the two-point correlator  explains why the velocity  $\vec{v}$ in the Li\-o\-uville operator, 
see equation \eqref{eq:liouvilleeq}, does not carry any generation indices. The velocity is determined by the 
value of $\epsilon$ on the respective shell and is different for the diagonal and off-diagonal components  
of $\varrho_{ij}(t,\vec{x},\epsilon,\vec{p})$. In the plane wave approximation the diagonals propagate with the velocity of the 
respective mass eigenstate, $\vec{p}_w/\omega_i(\vec{p}_w)$, whereas the off-diagonals propagate with the velocity of the 
center of momentum, $2\vec{p}_w/(\omega_i(\vec{p}_w)+\omega_j(\vec{p}_w))$.

\paragraph{Derivation from the Schr\"odinger equation.}
Because the dynamics of the underlying wave functions is go\-verned by the Schr\"odinger equation, one can expect that, 
similarly to the Liouville--von Neumann equation for the neutrino density matrix, the evolution equation for the neutrino 
two-point correlator \eqref{eq:liouvilleeq} can be derived from the Schr\"odinger equation without using an explicit solution 
of the latter. To demonstrate this we first recast equation \eqref{eq:liouvilleeq} in the form
\begin{align}
\label{eq:liouvilleeqtpc}
(\epsilon\partial_t+\vec{p}\partial_\vec{x})\varrho_{ij}(t,\vec{x},\epsilon,\vec{p}) =
 -{\textstyle\frac{i}{2}}(m^2_i-m^2_j)\varrho_{ij}(t,\vec{x},\epsilon,\vec{p})\,.
\end{align}
Next, we differentiate equation \eqref{eq:correlator} with respect to the central time $t$, 
multiply the result of differentiation by $\epsilon$, and further use 
$\epsilon e^{i\epsilon \tau}=-i\partial_{\tau} e^{i\epsilon \tau}$ on the right-hand side. Integrating by parts we can replace 
$-i\partial_{\tau} e^{i\epsilon \tau}$ by $i e^{i\epsilon \tau} \partial_{\tau}$. Using the chain rule we finally arrive at 
\begin{align}
\label{eq:liouvillevacstep1}
\epsilon\partial_t \varrho_{ij}(t,\vec{x},\epsilon,\vec{p}) &  = \frac{i}{2}
\int d\tau e^{i\epsilon\tau} \int d^3	\vec{y} e^{-i\vec{p}\vec{y}}  \nonumber\\ 
&\times [\partial^2_t\psi_i(t+\tau/2,\vec{x}+\vec{y}/2)\cdot \psi^{*}_j(t-\tau/2,\vec{x}-\vec{y}/2)\nonumber\\
&-\psi_i(t+\tau/2,\vec{x}+\vec{y}/2)\cdot \partial^2_t\psi^{*}_j(t-\tau/2,\vec{x}-\vec{y}/2)]\,.
\end{align}
Proceeding similarly, i.e. differentiating equation \eqref{eq:correlator} with respect to the central coordinate $\vec{x}$, 
mu\-ltiplying the result of differentiation by $\vec{p}$, using $e^{-i\vec{p}\vec{y}}\vec{p}=i\partial_\vec{y}e^{-i\vec{p}\vec{y}}$, 
integrating by parts to replace $i\partial_\vec{y}e^{-i\vec{p}\vec{y}}$ by $-ie^{-i\vec{p}\vec{y}}\partial_\vec{y}$, and finally 
using the chain rule we obtain   
\begin{align}
\label{eq:liouvillevacstep2}
\vec{p}\partial_\vec{x} \varrho_{ij}(t,\vec{x},\epsilon,\vec{p}) &  = -\frac{i}{2}
\int d\tau e^{i\epsilon\tau} \int d^3	\vec{y} e^{-i\vec{p}\vec{y}}  \nonumber\\ 
&\times [\partial^2_\vec{x}\psi_i(t+\tau/2,\vec{x}+\vec{y}/2)\cdot \psi^{*}_j(t-\tau/2,\vec{x}-\vec{y}/2)\nonumber\\
&-\psi_i(t+\tau/2,\vec{x}+\vec{y}/2)\cdot \partial^2_\vec{x}\psi^{*}_j(t-\tau/2,\vec{x}-\vec{y}/2)]\,.
\end{align}
Combining equations \eqref{eq:liouvillevacstep1} and \eqref{eq:liouvillevacstep2} we arrive at	 
\begin{align}
\label{eq:liouvillevacstep3}
(\epsilon\partial_t+\vec{p}\partial_\vec{x})\varrho_{ij}(t,\vec{x},\epsilon,\vec{p}) &  = \frac{i}{2}
\int d\tau e^{i\epsilon\tau} \int d^3	\vec{y} e^{-i\vec{p}\vec{y}}  \nonumber\\ 
&\times [\square\psi_i(t+\tau/2,\vec{x}+\vec{y}/2)\cdot \psi^{*}_j(t-\tau/2,\vec{x}-\vec{y}/2)\nonumber\\
&-\psi_i(t+\tau/2,\vec{x}+\vec{y}/2)\cdot \square\psi^{*}_j(t-\tau/2,\vec{x}-\vec{y}/2)]\,,
\end{align}
where $\square\equiv \partial^2_t-\partial^2_\vec{x}$ is the  d'Alembert operator. The Schr\"odinger equation 
\eqref{eq:schroedinger} implies that in vacuum $\partial^2_t \psi_i(t,\vec{x})=-{\sf K}^2_{ii}(\vec{x})\psi_i(t,\vec{x})
=(\partial^2_\vec{x}-m_i^2)\psi_i(t,\vec{x})$. Action of the  d'Alembert operator hence results in $\square\psi_i = 
-m^2_i\psi_i$, as expected. Comparing the right-hand side of equation \eqref{eq:liouvillevacstep3} to the definition 
of the two-point correlator we conclude that the former  reduces to 
$-\frac{i}{2}(m^2_i-m^2_j)\varrho_{ij}(t,\vec{x},\epsilon,\vec{p})$. In other words, we recover equation \eqref{eq:liouvilleeqtpc}. 

\paragraph{Quantum-kinetic equation.}
Equation \eqref{eq:liouvilleeqtpc} can be rewritten in a manifestly covariant form,
\begin{align}
\label{eq:liouvilleeqcov}
p^\mu \partial_{\mu}\varrho_{ij}(x,p) = -{\textstyle\frac{i}{2}}(m^2_i-m^2_j)\varrho_{ij}(x,p)\,,
\end{align}
where $x^\mu=(t,\vec{x})$ and $p^\mu=(\epsilon,\vec{p})$, see appendix \ref{sec:covliouville} for more details. Its form 
matches the form of the quantum kinetic equation for the left-handed neutrino component (considered in the vacuum limit)
discussed in reference \cite{Yamada:2000za}. In the quantum kinetic approach to  oscillations the two-point 
correlator $\varrho(x,p)$ is 
re\-placed by the Wightman propagator $\varrho_<(x,p)$, defined as the expectation value of the neutrino field operators
\citep{Yamada:2000za,Vlasenko:2013fja,Cirigliano:2014aoa, Vlasenko:2014bva,Blaschke:2016xxt,Richers:2019grc}. A 
few further details are presented in appendix \ref{sec:quantumkineticeq}.
Apart from the quantum kinetic equation, the Wightman propagator must also satisfy the complementary constraint equation 
that determines its spectrum.

In addition to the neutrino and antineutrino correlators, the formalism of the non-equilibrium quantum field 
theory also naturally incorporates the particle-antiparticle correlators \cite{Herranen:2008di,Herranen:2010mh,Herranen:2011zg,
Fidler:2011yq}. The authors of references \cite{Blasone:1998hf,Bernardini:2004wr,Blasone:2018ktu,Blasone:2019rxl} have shown
that the particle-antiparticle 
correlations also affect neutrino oscillations in vacuum, albeit their contribution is strongly suppressed in the 
physically interesting relativistic limit. As has been argued in reference \cite{Kobach:2017osm}, they are nevertheless
important to ensure that neutrino detection probability exactly vanishes outside of the light cone. An approach, in which 
faster-than-light signaling is automatically and manifestly prevented, has been developed in reference \cite{Dickinson:2016oiy}.
This approach works directly with probabilities instead of squared matrix elements. The probability-level calculation implicitly 
sums over the unobserved emissions that are crucial to ensuring that the measurement is local and consistent with causality. 
The need to sum inclusively over unobserved emissions was first described in reference \cite{Ferretti:1968}
and made explicit in reference \cite{PhysRevA.56.3395}.

\subsection{\label{sec:kinetics}Description in terms of the Wigner function}
The Wigner function $\varrho_{ij}(t,\vec{x},\vec{p})$ is obtained from the two-point correlator by integration over $\epsilon$. 
Since the Liouville term also 
depends on $\epsilon$, strictly speaking, the integration does not yield a closed-form evolution
equation for the Wigner function. 
On the other hand, for typical initial conditions the width of the two-point correlator in the $\epsilon$ space is negligibly small and 
the off-shell contributions can be neglected. In the relativistic limit this results in an evolution 
equation for the Wigner function whose form matches the form of (the vacuum limit of) the kinetic equation of Sigl and Raffelt.

The latter is  believed to be inconsistent with the quantum mechanical  uncertainty principle. We argue	 
below that  this is not the case. Applied to the Wigner function, the uncertainty principle 
states that the stronger the Wigner function is peaked around a characteristic momentum  $\vec{p}_w$, the weaker is its 
dependence on $\vec{v}t-\vec{x}$, and vice versa. If the Wigner function initially satisfies the uncertainty principle, the 
evolution equation ensures that this holds true in the course of the neutrino propagation. However, unlike the 
Schr\"odinger equation, the evolution equation itself does not enforce the inverse relation between localization 
in the coordinate and momentum spaces. 

Each individual momentum mode of the Wigner function does not experience any suppression in the course of the neutrino 
propagation. On the other hand, because of nonzero width of the Wigner function in the momentum space related to the 
uncertainty principle, off-diagonal components of the density matrix do get suppressed at late times due to
dephasing of the momentum modes.

\paragraph{Evolution equation for Wigner function.}
In the plane wave limit the Liouville equation \eqref{eq:liouvilleeq} can be trivially integrated over $\epsilon$. 
On the left-hand side the integration yields a matrix of velocities,
\begin{align}
\label{eq:propvelocity}
\vec{v}_{ij}(\vec{p})\equiv \frac{2\vec{p}}{\omega_i(\vec{p})+\omega_j(\vec{p})}\,.
\end{align}
On the right-hand side the integration yields the familiar 

\begin{align}
\label{eq:usualrhs}
\frac{m_i^2-m_j^2}{2\epsilon} \rightarrow \frac{m^2_i-m^2_j}{\omega_i(\vec{p})+\omega_j(\vec{p})} =
\frac{\omega^2_i(\vec{p})-\omega^2_j(\vec{p})}{\omega_i(\vec{p})+\omega_j(\vec{p})}=
\omega_i(\vec{p})-\omega_j(\vec{p})\,.
\end{align}
Let us emphasize that equation \eqref{eq:usualrhs} is a direct consequence of the fact that the off-diagonals of the 
two-point correlator live on the intermediate shell \cite{Kartavtsev:2015vto}. The resulting generalization of the classical 
Liouville equation to the case of mixing neutrinos reads
\begin{align}
\label{eq:liouvillewigner}
(\partial_t+\vec{v}_{ij}(\vec{p})\partial_\vec{x})\varrho_{ij}(t,\vec{x},\vec{p}) = 
-i(\omega_i(\vec{p})-\omega_j(\vec{p}))\varrho_{ij}(t,\vec{x},\vec{p})\,.
\end{align}
Also for initial conditions other than the plane wave ones, the two-point correlator is typically strongly peaked in the 
vicinity of the respective shells. In this case the small variations of $\epsilon$ around the on-shell value  can be safely 
neglected and integration of equation \eqref{eq:liouvilleeq} again yields equation \eqref{eq:liouvillewigner}.

\paragraph{The uncertainty principle.}
In the plane wave approximation the neutrino is completely localized in the momentum space and completely delocalized 
in the coordinate space. Let us now return to the case of Gaussian initial conditions, see equation \eqref{eq:gaussianwp}.

Trading the integration variables $\vec{k}$ and $\vec{q}$ in equation \eqref{eq:shapefactor} for the center and relative 
momentum, $\vec{s}\equiv (\vec{k}+\vec{q})/2$ and $\vec{\Delta}\equiv \vec{k}-\vec{q}$, and integrating over the center 
momentum we obtain for the shape factor 
\begin{align}
\label{eq:shapefactorrel}
g_{ij}(\vec{v}t-\vec{x},\epsilon,\vec{p})  & = \int \frac{d^3\vec{\Delta}}{(2\pi)^3} 
(2\pi)\,\delta\biggl(\epsilon-\frac{\omega_i(\vec{p}+{\textstyle\frac12}\vec{\Delta})
+\omega_j(\vec{p}-{\textstyle\frac12}\vec{\Delta})}2\biggr) 
\nonumber\\
& \times \psi_i(0,\vec{p}+{\textstyle\frac12}\vec{\Delta}) \psi^*_j(0,\vec{p}-{\textstyle\frac12}\vec{\Delta}) 
\,e^{-i\vec{\Delta}(\vec{v}t-\vec{x})}\,.
\end{align}
Because contributions of $\vec{\Delta}$ larger than the neutrino momentum uncertainty are strongly suppressed, for 
$\sigma_p \ll \vec{p}$ we can safely approximate $\omega_i(\vec{p}+\vec{\Delta}/2)+\omega_j(\vec{p}-\vec{\Delta}/2)$ by 
$\omega_i(\vec{p})+\omega_j(\vec{p})$, thereby approximating the propagation velocity by $\vec{v}_{ij}(\vec{p})$.
The integration over $\epsilon$ is then trivial. The resulting approximate Wigner function also factorizes into a shape and a 
phase factor,
\begin{align}
\label{eq:shapepahsewigner}
\varrho_{ij}(t,\vec{x},\vec{p}) \approx 
e^{-i\omega_i(\vec{p})t}\,g_{ij}(\vec{v}t-\vec{x},\vec{p})\,e^{i\omega_j(\vec{p})t}\,,
\end{align}
with the shape factor given by
\begin{align}
\label{eq:shapewigner}
g_{ij}(\vec{v}t-\vec{x},\vec{p})  & = \int \frac{d^3\vec{\Delta}}{(2\pi)^3} 
\, \psi_i(0,\vec{p}+{\textstyle\frac12}\vec{\Delta}) \psi^*_j(0,\vec{p}-{\textstyle\frac12}\vec{\Delta}) 
\,e^{-i\vec{\Delta}(\vec{v}_{ij}(\vec{p})t-\vec{x})}\,.
\end{align}
As can be verified by substitution, equation \eqref{eq:shapepahsewigner} satisfies the evolution equation 
\eqref{eq:liouvillewigner} for arbitrary initial conditions. For Gaussian initial conditions the remaining integration over 
$\vec{\Delta}$ in equation \eqref{eq:shapewigner} can be done analytically and yields
\begin{align}
\label{eq:shapewignergaussian}
g_{ij}&(\vec{v}t-\vec{x},\vec{p}) \propto \exp\biggl(-\frac{(\vec{p}-\vec{p}_w)^2}{2\sigma_p^2}\biggr)\,
\exp\biggl(-\frac{(\vec{v}_{ij}(\vec{p})t-\vec{x})^2}{2\sigma^2_x}\biggr)\,.
\end{align}
The form of the shape factor for Lorentzian initial conditions is presented in appendix \ref{sec:lorentzian}.

Equation \eqref{eq:shapewignergaussian} implies that the evolution equation admits solutions consistent with the 
uncertainty principle. Note, however, that equation \eqref{eq:shapepahsewigner} remains a solution of the evolution equation 
even if  $\sigma_p$ and $\sigma_x$ are considered as independent. In particular, one can simultaneously take the limits 
$\sigma_p\rightarrow 0$ and $\sigma_x \rightarrow 0$, thereby `creating' a state that is localized both in the momentum 
and coordinate spaces. That is, the evolution equation itself does not enforce the inverse relation between localization 
in the coordinate and momentum spaces. The latter is encoded in the initial conditions for $\varrho_{ij}(t,\vec{x},\vec{p})$
instead.

\paragraph{Kinematical decoherence.} 
In the  approximation used to derive equation \eqref{eq:shapepahsewigner}, individual momentum modes of the Wigner 
function do not experience any suppression in the course of the neutrino propagation. On the other hand, their growing 
dephasing results in kinematical decoherence and related suppression of the off-diagonals of the density matrix, 
known as wave packet separation \cite{Akhmedov:2009rb,Stirner:2018ojk}. 
 
To  demonstrate this we derive the respective density matrix by integrating equation \eqref{eq:shapepahsewigner} over 
$\vec{p}$. For a wave function strongly peaked around the characteristic momentum $\vec{p}_w$ the Wigner  
function can be expanded in the vicinity of $\vec{p}_w$. In the relativistic limit 
$\vec{v}_{ij}(\vec{p})\sim \vec{1}$ and is not sensitive to small, of the order of $\sim \sigma_p$, variations of 
$\vec{p}$ around $\vec{p}_w$. Therefore, we can  approximate $\vec{v}_{ij}(\vec{p})$ by $\vec{v}_{ij}(\vec{p}_w)$
in the shape factor, see equation \eqref{eq:shapewigner}. On the other hand, expanding the argument of the phase factor, 
$\omega_i(\vec{p})-\omega_j(\vec{p})\approx \omega_i(\vec{p}_w)-\omega_j(\vec{p}_w)+\Delta\vec{v}_{ij}(\vec{p}_w)
(\vec{p}-\vec{p}_w)$, we find that it is sensitive to the difference of the respective group velocities. In this approximation
the Wigner function reads
\begin{align}
\label{eq:wignerapprox}
\varrho_{ij}(t,\vec{x},\vec{p}) \approx \varrho_{ij}(t,\vec{x},\vec{p}_w)\cdot
\exp\bigl(-(\vec{p}-\vec{p}_w)^2/2\sigma_p^2-i\Delta\vec{v}_{ij}(\vec{p}_w)t\,(\vec{p}-\vec{p}_w)\bigr)\,.
\end{align}
Integrating over $\vec{p}$ we recover equation \eqref{eq:densitymatrixgaussian},
\begin{align}
\label{eq:densmatrwignerapprox} 
\rho_{ij}(t,\vec{x}) \approx \varrho_{ij}(t,\vec{x},\vec{p}_w)\cdot 
\exp\biggl(-\frac{\Delta\vec{v}^2_{ij}(\vec{p}_w)t^2}{8\sigma^2_x}\biggr)\,.
\end{align}
Therefore, kinematical decoherence in the density matrix is caused by dephasing of the individual momentum modes, which 
grows in the course of the neutrino  propagation. This picture is equivalent to the one of the wave packet separation 
\cite{Akhmedov:2009rb}. The smaller is the considered momentum range, the less pronounced is the suppression. 
This corresponds to coherence restoration in a detector with a very good energy resolution. Let us emphasize, that the finite 
energy resolution of a detector and an emission process can just as well be taken into account by post-processing the results 
through an integral over energy, instead of considering solutions of finite width in the momentum space.

The nonzero width of the  wave packets in the momentum space, crucial for the onset of kinematical decoherence at late 
times,  can be traced back to the uncertainty principle and a nonzero spatial size of the wave packets \cite{Akhmedov:2017mcc}. 
It is interesting to note, however, that if we were to treat $\sigma_p$ and $\sigma_x$ as independent and set the latter to zero in 
equation \eqref{eq:shapewigner}, we would still get the same suppression factor in  equation \eqref{eq:densmatrwignerapprox}. 
This (inconsistent from the viewpoint of quantum mechanics) approximation still solves equation \eqref{eq:liouvillewigner},
as has been discussed above. If we considered an ensemble of neutrinos, this approximation would correspond to describing 
neutrinos as classical particles with a particular momentum distribution. This picture has been used in reference \cite{Raffelt:2010za}   
to study collective oscillations of a two-flavor neutrino system. Hence, the results obtained there do take into account the effect of 
wave packet separation, at least for low neutrino densities. For recent developments in the field of collective neutrino oscillations 
see e.g. references \cite{Airen:2018nvp,Capozzi:2018clo,Capozzi:2019lso,Abbar:2019zoq,Johns:2019izj,Chakraborty:2019wxe,Cervia:2019res,Doring:2019axc,Shalgar:2019qwg}.

As has been noted  in references \cite{Hansen:2016klk,Akhmedov:2017mcc}, because $\varrho_{ij}(t,\vec{x},\vec{p}_w)$
satisfies the evolution equation \eqref{eq:liouvillewigner} with $\vec{p}=\vec{p}_w$ the density matrix also satisfies 
an effective Liouville equation,
\begin{align}
\label{eq:effeqfordensmatr}
(\partial_t+\vec{v}_{ij}(\vec{p}_w)\partial_\vec{x})\rho_{ij}(t,\vec{x}) \approx 
-i(\omega_i(\vec{p}_w)-\omega_j(\vec{p}_w))\rho_{ij}(t,\vec{x}) + {\cal C}_{ij}(t,\vec{x})\,,
\end{align}
where ${\cal C}_{ij}(t,\vec{x})$ stems from the last term on the right-hand side of equation \eqref{eq:densmatrwignerapprox} 
and describes kinematical decoherence. The explicit form of ${\cal C}(t,\vec{x})$ depends on the initial conditions and can be 
calculated a posteriori, once the solution for the density matrix is known \cite{Hansen:2016klk}.

\paragraph{Spread of the wave packet.}
Because different momentum modes propagate with different velocities, the width of the density matrix increases with time. 
To estimate this effect we expand $\vec{v}_{ij}(\vec{p})$ in equation \eqref{eq:shapewigner} in the vicinity of $\vec{p}_w$, 
$\vec{v}_{ij}(\vec{p})\approx \vec{v}_{ij}(\vec{p}_w)+\vec{v}'_{ij}(\vec{p}_w)(\vec{p}-\vec{p}_w)$ (note that the expansion 
is understood to be made in the direction of the neutrino propagation). Integrating over $\vec{p}$ we recover equation 
\eqref{eq:densitymatrixgaussian} with $\sigma^2_x$ replaced by an effective  generation-  and time-dependent width
\cite{Kersten:2015kio},
\begin{equation}
\sigma^2_{x,{\rm eff}}\approx \sigma^2_x + (\sigma_p\vec{v}'_{ij}(\vec{p}_w)	t)^2\,.
\end{equation}
The phase factor also receives a tiny, of the order of $\Delta \vec{v}_{ij}(\vec{p}_w)\vec{v}'_{ij}(\vec{p}_w)$, correction
that is proportional to $\vec{v}_{ij}(\vec{p}_w)t-\vec{x}$ and hence vanishes at the center of the wave packet. 

\paragraph{Off-shell effects.} 
As can be read off from equation \eqref{eq:shapefactorrel}, for each $\vec{p}$ there is a  range of allowed values of 
$\epsilon$. In other words, the two-point correlator possesses an off-shell component. Therefore, even for a fixed value 
of the momentum, $\varrho_{ij}(t,\vec{x},\vec{p})$ is a superposition of plane waves propagating with different velocities 
$\vec{v}=\vec{p}/\epsilon$ centered around $\vec{v}_{ij}$ with a tiny spread
$\sim \vec{v}_{ij}(\vec{p}) \Delta \vec{v}_{ij}(\vec{p}) \cdot \sigma_p/(\omega_i(\vec{p})+\omega_j(\vec{p}))$. 
By analogy with the wave packet separation one could expect kinematical decoherence in the Wigner function caused by 
a gradual dephasing of the individual modes at late times. Analysis of this subtle effect is beyond the scope of the present 
work.

\paragraph{Kinetic equation.}
In the relativistic limit $\vec{v}_{ij}$ is well approximated by the average velocity $\bar{\vec{v}}_{ij}$ in\-troduced earlier. In this 
approximation the evolution equation \eqref{eq:liouvillewigner} can be recast in the form that matches the form of the kinetic 
equation discussed in references \cite{Sigl:1992fn, Stirner:2018ojk},
\begin{align}
\label{eq:liouvillewigneraltrarel}
\partial_t\varrho(t,\vec{x},\vec{p})
+{\textstyle\frac12}\{\partial_\vec{p}\omega,\partial_\vec{x}\varrho(t,\vec{x},\vec{p})\}
=-i\,[\omega(\vec{p}),\varrho(t,\vec{x},\vec{p})]\,,
\end{align} 
where $[.\,,.]$ denotes a commutator, and $\{.\,,.\}$ an anticommutator.  In the kinetic equation
$\varrho(t,\vec{x},\vec{p})$ is replaced by the matrix of densities, defined as the expectation value of bilinears 
of the creation and annihilation operators of the neutrino field \cite{Sigl:1992fn}.

\section{\label{sec:quantumkinetics}Neutrino propagation in matter}

In this section we generalize the results of section \ref{sec:liouvilleterm} to neutrino propagation in an external potential. In 
subsection \ref{sec:quantumkineticsmed} we derive, to the first order in the gradient expansion, an evolution
equation for the neutrino two-point correlator, valid also for non-relativistic neutrinos. Up to small corrections related to a 
somewhat different expansion scheme its form matches (the collisionless limit of) the quantum kinetic equation. Analyzing 
the energy dependence of its terms we derive a simplified evolution equation valid in the relativistic limit. 

In subsection \ref{sec:kineticsmed} we use the latter to derive an evolution equation for the Wigner function.
Its form matches (the collisionless limit of) the kinetic equation of Sigl and Raffelt. Considering Gaussian initial conditions we 
show that it admits solutions consistent with the uncertainty principle and accounts for neutrino wave packet separation. We 
limit our analysis to forward scattering and do not consider the Lindblad terms 
\cite{Lindblad:1975ef,Benatti:2000ph,PhysRevA.45.2243} describing dynamical decoherence.

\subsection{\label{sec:quantumkineticsmed}Description in terms of the two-point correlator}

Because the Liouville operator on the left-hand side of the evolution equation for the two-point correlator,
$\epsilon\partial_t + \vec{p} \partial_\vec{x}$, 
has the dimension of energy squared, its right-hand side contains (anti) commutators of the two-point cor\-relator with the 
Hamiltonian squared. For non-commuting kinetic and potential energy operators this yields a new contribution proportional to 
their commutator. Shifts of the on-shell value of $\epsilon$ induced by a time-dependent potential also result in a new contribution
proportional to the time-derivative of the potential. In the relativistic limit, which we consider first to make contact with results of 
references \cite{Sigl:1992fn,Stirner:2018ojk}, these new terms can be neglected and the evolution equation 
reverts to the conventional one.

\paragraph{Relativistic limit.} To derive evolution equation for the two-point correlator we differentiate equation
\eqref{eq:correlator} with respect to time. The time derivatives of the wave functions in the integrand of the resulting 
expression are determined  by the Schr\"odinger equation \eqref{eq:schroedinger},
\begin{align}
\label{eq:rellimitstep1}
\partial_t \varrho_{ij}(t,\vec{x},\epsilon,\vec{p}) &  =
-i\int d\tau e^{i\epsilon\tau} \int d^3	\vec{y} e^{-i\vec{p}\vec{y}}  \nonumber\\ 
&\times [{\sf H}_{in}(t+\tau/2,\vec{x}+\vec{y}/2) \psi_n(t+\tau/2,\vec{x}+\vec{y}/2)\cdot \psi^{*}_j(t-\tau/2,\vec{x}-\vec{y}/2)\nonumber\\
&-\psi_i(t+\tau/2,\vec{x}+\vec{y}/2)\cdot {\sf H}_{jm}(t-\tau/2,\vec{x}-\vec{y}/2)\psi^{*}_m(t-\tau/2,\vec{x}-\vec{y}/2)]\,.
\end{align}
In order to keep the derivation as simple as possible we assume that the background matter is unpolarized and 
also free of convective currents. With these assumption the matter is characterized by a potential ${\sf V}_{ij}(t,\vec{x})$
determined by occupation numbers of the charged leptons.
Because  equation \eqref{eq:rellimitstep1} is linear in the Hamiltonian, ${\sf H}_{ij}(t,\vec{x}) ={\sf K}_{ij}(\vec{x})+
{\sf V}_{ij}(t,\vec{x})$, contributions from the kinetic and potential terms can be treated separately. 

The contribution of the potential term can be conveniently analyzed in the coordinate representation employed in equations 
\eqref{eq:correlator} and \eqref{eq:rellimitstep1}.  Expanding ${\sf V}_{ij}(t\pm\tau/2,\vec{x}\pm\vec{y}/2)$ to the first 
order in $\tau$ and $\vec{y}$,  using $e^{i\epsilon\tau}\tau = -i\partial_\epsilon e^{i\epsilon\tau}$ and 
$e^{-i\vec{p}\vec{y}}\vec{y}=i\partial_\vec{p} e^{-i\vec{p}\vec{y}}$ in equation \eqref{eq:rellimitstep1}, and comparing 
the resulting expression to the definition of the two-point correlator, see equation \eqref{eq:correlator}, we find for contribution
of the potential term 
\begin{align}
\label{eq:ultrarelapproxpot}
\partial_t \varrho(t,\vec{x},\epsilon,\vec{p}) \ni  &
+ {\textstyle\frac12}\{\partial_\vec{x} {\sf V}(t,\vec{x}),\partial_\vec{p}\varrho(t,\vec{x},\epsilon,\vec{p})\}
-{\textstyle\frac12}\{\partial_t {\sf V}(t,\vec{x}),\partial_\epsilon\varrho(t,\vec{x},\epsilon,\vec{p})\}\nonumber\\
&-i\,[{\sf V}(t,\vec{x}),\varrho(t,\vec{x},\epsilon,\vec{p})]\,.
\end{align}
Contribution of the kinetic term is more easily analyzed in the momentum representation where the action of the kinetic 
operator amounts to multiplication of the wave functions  by $\omega$,
\begin{align}
\partial_t\varrho_{ij}(t,\vec{x},\epsilon,\vec{p}) \ni & -i
\int d\tau e^{i\epsilon\tau} \int d^3	\vec{y} e^{-i\vec{p}\vec{y}} 
\int \frac{d^3\vec{s}}{(2\pi)^3} \frac{d^3\vec{\Delta}}{(2\pi)^3}e^{i\vec{\Delta}\vec{x}}e^{i\vec{s}\vec{y}}\nonumber\\ 
&\times [\omega_i(\vec{s}+\vec{\Delta}/2)\psi_i(t+\tau/2,\vec{s}+\vec{\Delta}/2)\cdot \psi^{*}_j(t-\tau/2,\vec{s}-\vec{\Delta}/2)\nonumber\\
&-\psi_i(t+\tau/2,\vec{s}+\vec{\Delta}/2)\cdot \omega_j(\vec{s}-\vec{\Delta}/2) \psi^{*}_j(t-\tau/2,\vec{s}-\vec{\Delta}/2)]\,.
\end{align}
To leading order in the gradients $\omega_i(\vec{p}+\vec{\Delta}/2)-\omega_j(\vec{p}-\vec{\Delta}/2) \approx 
\bar{\vec{v}}_{ij}(\vec{p})\vec{\Delta}+\omega_i(\vec{p})-\omega_j(\vec{p})$ in the relativistic limit. Using 
$e^{i\vec{\Delta}\vec{x}} \vec{\Delta}=-i\partial_\vec{x} e^{i\vec{\Delta}\vec{x}}$ and comparing the resulting 
expression to the definiti\-on of the two-point correlator in the momentum representation, see appendix 
\ref{sec:constantpot}, we find that in the relativistic limit contribution of the kinetic term can be approximated by 
\begin{align}
\label{eq:ultrarelapproxkin} 
\partial_t\varrho(t,\vec{x},\epsilon,\vec{p}) \ni 
-{\textstyle\frac12}\{\partial_\vec{p}\omega(\vec{p}),\partial_\vec{x}\varrho(t,\vec{x},\epsilon,\vec{p})\}
-i[\omega(\vec{p}),\varrho(t,\vec{x},\epsilon,\vec{p})]\,.
\end{align}
Introducing ${\sf H}_{ij}(t,\vec{x},\vec{p})\equiv \delta_{ij}\omega_j(\vec{p})+{\sf V}_{ij}(t,\vec{x})$ we can combine equations 
\eqref{eq:ultrarelapproxpot}  and \eqref{eq:ultrarelapproxkin} to 
\begin{align}
\label{eq:ultrarelapprox}
\partial_t \varrho(t,\vec{x},\epsilon,\vec{p})  
&+{\textstyle\frac12}\{\partial_\vec{p} {\sf H}(t,\vec{x},\vec{p}),\partial_\vec{x}\varrho(t,\vec{x},\epsilon,\vec{p})\}
-{\textstyle\frac12}\{\partial_\vec{x} {\sf H}(t,\vec{x},\vec{p}),\partial_\vec{p}\varrho(t,\vec{x},\epsilon,\vec{p})\}
\nonumber\\
&+{\textstyle\frac12}\{\partial_t {\sf H}(t,\vec{x},\vec{p}),\partial_\epsilon\varrho(t,\vec{x},\epsilon,\vec{p})\}
\approx -i\,[{\sf H}(t,\vec{x},\vec{p}),\varrho(t,\vec{x},\epsilon,\vec{p})]\,,
\end{align}
valid in the relativistic limit. The first two terms on the left-hand side generalize the flux conservation equation. The third 
one describes momentum changes by coherent external forces \cite{Stirner:2018ojk}. Finally, the fourth one describes  
spectrum shifts caused by a changing external potential. 

As we have seen in section \ref{sec:liouvilleterm}, in order to derive a closed-form evolution equation without resorting 
to the relativistic limit one has to work with the Liouville operator of the form $\epsilon\partial_t + \vec{p}\partial_\vec{x}$. 
The form of the resulting evolution equation in matter is discussed in the remainder of this subsection.  

\paragraph{Neutrino propagation in a constant potential.}
Before generalizing equation \eqref{eq:liouvilleeqcov} to space- or time-dependent potentials let us first consider 
the technically simple case of a constant potential ${\sf V}_{ij}$. By analogy with equation \eqref{eq:liouvilleeqcov} we multiply 
equ\-ation \eqref{eq:rellimitstep1} by $\epsilon$ and further use $\epsilon e^{i\epsilon \tau}=-i\partial_{\tau} e^{i\epsilon \tau}$ 
on the right-hand side. Integrating by parts we can replace $-i\partial_{\tau} e^{i\epsilon \tau}$ by 
$i e^{i\epsilon \tau} \partial_{\tau}$. Because the Ha\-miltonian ${\sf H}_{ij}(\vec{x})={\sf K}_{ij}(\vec{x})+{\sf V}_{ij}$ is 
time-independent, subsequently using the chain rule and the Schr\"odinger equation \eqref{eq:schroedinger}  we arrive at
\begin{align}
\label{eq:qkconstpotstep1}
\epsilon\,\partial_t \varrho_{ij}(t,\vec{x},\epsilon,\vec{p}) &  =
-\frac{i}2\int d\tau e^{i\epsilon\tau} \int d\vec{y} e^{-i\vec{p}\vec{y}}  \nonumber\\ 
&\times [{\sf H}^2_{in}(\vec{x}+\vec{y}/2) \psi_n(t+\tau/2,\vec{x}+\vec{y}/2)\cdot \psi^{*}_j(t-\tau/2,\vec{x}-\vec{y}/2)\nonumber\\
&-\psi_i(t+\tau/2,\vec{x}+\vec{y}/2)\cdot {\sf H}^2_{jm}(\vec{x}-\vec{y}/2)\psi^{*}_m(t-\tau/2,\vec{x}-\vec{y}/2)]\,,
\end{align}
where  we employ a compact notation ${\sf H}^2_{ij}\equiv {\sf H}_{in}{\sf H}_{nj}$. In the momentum representation
action of the  Hamiltonian ${\sf H}_{ij}(\vec{x})={\sf K}_{ij}(\vec{x})+{\sf V}_{ij}$  amounts to multiplication 
of the wave function by ${\sf H}_{ij}(\vec{p})=\delta_{ij}\,\omega_i(\vec{p})+{\sf V}_{ij}$, see appendix \ref{sec:constantpot} 
for more details. Expressing  the coordinate-representation wave functions in equation \eqref{eq:qkconstpotstep1}  in terms 
of their momentum-representation counterparts and subsequently introducing the center and relative momenta we arrive at   
\begin{align}
\label{eq:qkconstpotstep2}
\epsilon\,\partial_t \varrho_{ij}(t,\vec{x},\epsilon,\vec{p}) &  = -\frac{i}2\int d\tau e^{i\epsilon\tau} 
 \int d^3	\vec{y} e^{-i\vec{p}\vec{y}} 
\int \frac{d^3\vec{s}}{(2\pi)^3} \frac{d^3\vec{\Delta}}{(2\pi)^3}e^{i\vec{\Delta}\vec{x}}e^{i\vec{s}\vec{y}}\nonumber\\ 
&\times \bigl[{\sf H}^2_{in}(\vec{s}+\vec{\Delta}/2)\, \psi_n(t+\tau/2,\vec{s}+\vec{\Delta}/2)
\psi^{*}_j(t-\tau/2,\vec{s}-\vec{\Delta}/2) \nonumber\\
& - \psi_i(t+\tau/2,\vec{s}+\vec{\Delta}/2)\psi^{*}_m(t-\tau/2,\vec{s}-\vec{\Delta}/2){\sf H}^2_{mj}(\vec{s}-\vec{\Delta}/2)\bigr]\,.
\end{align}
To the first order in the gradient expansion
${\sf H}^2_{ij}(\vec{s}\pm\vec{\Delta}/2)\approx {\sf H}^2_{ij}(\vec{s})\pm 
\textstyle{\frac12} \partial_{\vec{s}}{\sf H}^2_{ij}(\vec{s})	\vec{\Delta}$. 
Using   $\vec{\Delta}e^{i\vec{\Delta}\vec{x}}=-i\partial_{\vec{x}}e^{i\vec{\Delta}\vec{x}}$ we finally arrive at 
\begin{align}
\label{eq:liouvilleeqconstpot}
\epsilon\,\partial_t\varrho(t,\vec{x},\epsilon,\vec{p}) 
+ {\textstyle\frac14}\bigl\{\partial_{\vec{p}}{\sf H}^2(\vec{p}),\partial_{\vec{x}}\varrho(t,\vec{x},\epsilon,\vec{p})\bigr\}
\approx -{\textstyle\frac{i}{2}}\bigl[{\sf H}^2(\vec{p}),\varrho(t,\vec{x},\epsilon,\vec{p})\bigr]\,.
\end{align}
In vacuum ${\sf H}^2_{ij}(\vec{p})=\delta_{ij}\,\omega^2_j(\vec{p})$, $\partial_\vec{p}{\sf H}^2_{ij}(\vec{p})=\vec{p}$  and equation \eqref{eq:liouvilleeqconstpot} reverts to 
equation \eqref{eq:liouvilleeqcov}.  

In appendix \ref{sec:constantpot} equation \eqref{eq:liouvilleeqconstpot} is re-derived using the solution of the 
Schr\"odinger equation in a constant potential. As can be read off  from the explicit expression for the two-point correlator, 
its spectrum is determined by eigenvalues of the Hamiltonian and is shifted with respect to the vacuum spectrum by terms 
proportional to the potential. Thus, the evolution equation \eqref{eq:liouvilleeqconstpot} must be supplemented 
by  initial conditions that consistently account for modifications of the neutrino spectrum in an external potential. 
As has been mentioned above, in the quantum kinetic approach to neutrino oscillations the spectrum can be obtained from the
supplementary constraint equation.

\paragraph{Neutrino propagation in a spatially inhomogeneous potential.}
For a spatially inhomogeneous but  time-independent potential equation \eqref{eq:qkconstpotstep1} still holds. However, 
because the operators ${\sf K}_{ij}(\vec{x})$ and ${\sf V}_{ij}(\vec{x})$ no longer commute, it does not reduce to equation 
\eqref{eq:qkconstpotstep2} in the momentum representation. 
To compute the right-hand side of equation \eqref{eq:qkconstpotstep1} we reorder the operators in ${\sf H}^2$ such that 
${\sf V}$ is left to ${\sf K}$, plus a commutator,  
\begin{align}
\label{eq:Hsqordered}
{\sf H}^2_{ij}(\vec{x}) & = {\sf K}^2_{ij}(\vec{x}) + {\sf V}_{in}(\vec{x}) {\sf K}_{nj}(\vec{x}) + 
{\sf V}_{nj}(\vec{x}) {\sf K}_{in}(\vec{x}) +{\sf V}^2_{ij}(\vec{x})\nonumber\\
& + [{\sf K}_{in}(\vec{x}){\sf V}_{nj}(\vec{x})-{\sf V}_{nj}(\vec{x}){\sf K}_{in}(\vec{x})]\,.
\end{align}
Expanding the kinetic energy operator in powers of  $\partial_{\vec{x}}/m$ we find to the first order in the gradients  
\begin{align}
\label{eq:KVcommutator}
{\sf K}_{in}(\vec{x}){\sf V}_{nj}(\vec{x})-{\sf V}_{nj}(\vec{x}){\sf K}_{in}(\vec{x})\approx 
-i\delta_{in}\partial_\vec{x}V_{nj}(\vec{x})\cdot\sum_{\ell=0}^\infty a_\ell\,2\ell \frac{(-i\partial_\vec{x})^{2\ell-1}}{m_n^{2\ell-1}}\,,
\end{align}
where $a_\ell$ are coefficients of the Taylor expansion. 

Substituting the first line of equation \eqref{eq:Hsqordered} into equation \eqref{eq:qkconstpotstep1} and  trading the  
momentum integration variables for the center and relative momenta we obtain  
\begin{align}
\label{eq:qkinhompotstep1}
\epsilon\,\partial_t \varrho_{ij}(t,\vec{x},&\epsilon,\vec{p})   = -\frac{i}2\int d\tau e^{i\epsilon\tau} 
 \int d^3	\vec{y} e^{-i\vec{p}\vec{y}} 
\int \frac{d^3\vec{s}}{(2\pi)^3} \frac{d^3\vec{\Delta}}{(2\pi)^3}e^{i\vec{\Delta}\vec{x}}e^{i\vec{s}\vec{y}}\nonumber\\ 
&\times \bigl[{\sf H}^2_{in}(\vec{x}+\vec{y}/2,\vec{s}+\vec{\Delta}/2)\, \psi_n(t+\tau/2,\vec{s}+\vec{\Delta}/2)
\psi^{*}_j(t-\tau/2,\vec{s}-\vec{\Delta}/2) \nonumber\\
& - \psi_i(t+\tau/2,\vec{s}+\vec{\Delta}/2)\psi^{*}_m(t-\tau/2,\vec{s}-\vec{\Delta}/2)
{\sf H}^2_{mj}(\vec{x}-\vec{y}/2,\vec{s}-\vec{\Delta}/2)\bigr]\,,
\end{align}
where ${\sf H}_{ij}(\vec{x},\vec{p}) = \delta_{ij}\omega_j(\vec{p})+{\sf V}_{ij}(\vec{x})$. Next, we expand the square of
the Hamiltonian to the first order in the gradients, ${\sf H}^2(\vec{x}\pm\vec{y}/2,\vec{s}\pm\vec{\Delta}/2)
\approx {\sf H}^2(\vec{x},\vec{s}) +\partial_\vec{s} {\sf H}^2(\vec{x},\vec{s}) \,\vec{\Delta}/2
+\partial_\vec{x} {\sf H}^2(\vec{x},\vec{s}) \,\vec{y}/2$. The first term of the expansion yields the commutator
\begin{align}
\label{eq:qkinhompotstep2}
\epsilon\,\partial\varrho(t,\vec{x},\epsilon,\vec{p})\ni -{\textstyle\frac{i}{2}}[{\sf H}^2(\vec{x},\vec{p}),\varrho(t,\vec{x},\epsilon,\vec{p})]\,,
\end{align}
which generalizes the right-hand side of equation \eqref{eq:liouvilleeqconstpot} to spatially inhomogeneous potentials. 
Using $e^{i\vec{\Delta}\vec{x}}\vec{\Delta}=-i\partial_\vec{x}e^{i\vec{\Delta}\vec{x}}$ and the product rule we obtain for 
contribution of the second expansion term 
\begin{align}
\label{eq:qkinhompotstep3}
\epsilon\,\partial\varrho(t,\vec{x},\epsilon,\vec{p})&\ni 
-{\textstyle\frac14}\partial_\vec{x}\{\partial_\vec{p}{\sf H}^2(\vec{x},\vec{p}),\varrho(t,\vec{x},\epsilon,\vec{p})\}
+{\textstyle\frac14}\{\partial_\vec{x}\partial_\vec{p}{\sf H}^2(\vec{x},\vec{p}),\varrho(t,\vec{x},\epsilon,\vec{p})\}\nonumber\\
&=-{\textstyle\frac14}\{\partial_\vec{p}{\sf H}^2(\vec{x},\vec{p}),\partial_\vec{x}\varrho(t,\vec{x},\epsilon,\vec{p})\}\,,
\end{align}
which generalizes the second term on the left-hand side of equation \eqref{eq:liouvilleeqconstpot} to  spatially inhomogeneous
potentials. 
Using $e^{-i\vec{p}\vec{y}} \vec{y}=i\partial_\vec{p}e^{-i\vec{p}\vec{y}}$ we find for contribution of the third expansion term   
\begin{align}
\label{eq:qkinhompotstep4}
\epsilon\,\partial\varrho(t,\vec{x},\epsilon,\vec{p})&\ni {\textstyle\frac14}\partial_\vec{p}\{\partial_\vec{x}{\sf H}^2(\vec{x},\vec{p}),
\varrho(t,\vec{x},\epsilon,\vec{p})\}
={\textstyle\frac14}\{\partial_\vec{x}{\sf H}^2(\vec{x},\vec{p}),\partial_\vec{p}\varrho(t,\vec{x},\epsilon,\vec{p})\}\nonumber\\
& +{\textstyle\frac14}\{
\partial_\vec{p}\omega(\vec{p})\partial_\vec{x}{\sf V}(\vec{x})+\partial_\vec{x}{\sf V}(\vec{x})\partial_\vec{p}\omega(\vec{p}),
\varrho(t,\vec{x},\epsilon,\vec{p})\}\,.
\end{align}
Substituting the second line of equation \eqref{eq:Hsqordered} into equation \eqref{eq:qkconstpotstep1}, using the 
approximation equation \eqref{eq:KVcommutator}, and trading the  momentum integration variables for the center 
and relative momenta we find:
\begin{align}
\label{eq:qkinhompotstep5}
\epsilon\,\partial_t \varrho_{ij}&(t,\vec{x},\epsilon,\vec{p})   \ni -\frac12\int d\tau e^{i\epsilon\tau} 
 \int d^3	\vec{y} e^{-i\vec{p}\vec{y}} 
\int \frac{d^3\vec{s}}{(2\pi)^3} \frac{d^3\vec{\Delta}}{(2\pi)^3}e^{i\vec{\Delta}\vec{x}}e^{i\vec{s}\vec{y}}\nonumber\\ 
&\times [\partial_\vec{s}\omega_i(\vec{s}+\vec{\Delta}/2)\partial_\vec{x}{\sf V}_{in}(\vec{x}+\vec{y}/2)
\, \psi_n(t+\tau/2,\vec{s}+\vec{\Delta}/2)
\psi^{*}_j(t-\tau/2,\vec{s}-\vec{\Delta}/2) \nonumber\\
& + \psi_i(t+\tau/2,\vec{s}+\vec{\Delta}/2)\psi^{*}_m(t-\tau/2,\vec{s}-\vec{\Delta}/2)
\partial_\vec{x}{\sf V}_{mj}(\vec{x}-\vec{y}/2)\partial_\vec{s}\omega_j(\vec{s}-\vec{\Delta}/2)]\,.
\end{align}
Because the right-hand side of equation \eqref{eq:qkinhompotstep5} already contains a derivative of the potential, to the 
first order in the gradient expansion it is sufficient to approximate $\partial_\vec{s}\omega_i(\vec{s}\pm\vec{\Delta}/2)
\partial_\vec{x}{\sf V}_{in}(\vec{x}\pm\vec{y}/2)$ by $\partial_\vec{s}\omega_i(\vec{s})\partial_\vec{x}{\sf V}_{in}(\vec{x})$. 
In this approximation we obtain 
\begin{align}
\label{eq:qkinhompotstep6}
\epsilon\,\partial_t \varrho&(t,\vec{x},\epsilon,\vec{p}) \ni -{\textstyle \frac12} 
(\partial_\vec{p}\omega(\vec{p})\partial_\vec{x}{\sf V}(\vec{x})\varrho(t,\vec{x},\epsilon,\vec{p})
+\varrho(t,\vec{x},\epsilon,\vec{p})\partial_\vec{x}{\sf V}(\vec{x})\partial_\vec{p}\omega(\vec{p}))\,,
\end{align}
whose structure is similar to that of the second term on the right-hand side of equation \eqref{eq:qkinhompotstep4}.

Combining equations \eqref{eq:qkinhompotstep2}, \eqref{eq:qkinhompotstep3}, \eqref{eq:qkinhompotstep4} and 
\eqref{eq:qkinhompotstep6} we finally arrive at 
\begin{align}
\label{eq:qkeqinhompot}
\epsilon\,\partial_t \varrho(t,\vec{x},\epsilon,\vec{p}) & +
{\textstyle\frac14}\{\partial_\vec{p}{\sf H}^2(\vec{x},\vec{p}),\partial_\vec{x}\varrho(t,\vec{x},\epsilon,\vec{p})\}
-{\textstyle\frac14}\{\partial_\vec{x}{\sf H}^2(\vec{x},\vec{p}),\partial_\vec{p}\varrho(t,\vec{x},\epsilon,\vec{p})\}\nonumber\\
&\approx -{\textstyle\frac{i}{2}}[{\sf H}^2(\vec{x},\vec{p})+{\textstyle \frac{i}2}[\partial_\vec{x}{\sf V}(\vec{x}),\partial_\vec{p}
\omega(\vec{p})],\varrho(t,\vec{x},\epsilon,\vec{p})]\,.
\end{align}
Equation \eqref{eq:qkeqinhompot} implies that in a spatially inhomogeneous potential neutrino oscillations are governed by  
${\sf H}^2(\vec{x},\vec{p})+\frac{i}{2}[\partial_\vec{x}{\sf V}(\vec{x}),\partial_\vec{p}\omega(\vec{p})]$. As this term 
is a sum of the  Hamiltonian and the commutator of the kinetic and potential energy operators, its structure replicates 
the structure of equation \eqref{eq:Hsqordered}. 
 
\paragraph{Neutrino propagation in a time-dependent potential.} To generalize equation \eqref{eq:qkconstpotstep1} to 
homogeneous but time-dependent potentials we again multiply equation \eqref{eq:rellimitstep1} by $\epsilon$ and use 
$\epsilon e^{i\epsilon \tau}=-i\partial_{\tau} e^{i\epsilon \tau}$ on the right-hand side, subsequently replacing 
$-i\partial_{\tau} e^{i\epsilon \tau}$ by $i e^{i\epsilon \tau} \partial_{\tau}$ upon integration by parts. Because the Hamiltonian 
${\sf H}_{ij}(t,\vec{x})={\sf K}_{ij}(\vec{x})+{\sf V}_{ij}(t)$ is time-dependent, using the chain rule and the Schr\"odinger equation 
\eqref{eq:schroedinger}  we find that in addition to an expression similar to the right-hand side of equation \eqref{eq:qkconstpotstep1},
\begin{align}
\label{eq:qktimedeppotstep1} 
\epsilon\,\partial_t \varrho_{ij}&(t,\vec{x},\epsilon,\vec{p})  =
-\frac{i}2\int d\tau e^{i\epsilon\tau} \int d\vec{y} e^{-i\vec{p}\vec{y}}  \nonumber\\ 
&\times [{\sf H}^2_{in}(t+\tau/2,\vec{x}+\vec{y}/2) \psi_n(t+\tau/2,\vec{x}+\vec{y}/2)\cdot \psi^{*}_j(t-\tau/2,\vec{x}-\vec{y}/2)\nonumber\\
&-\psi_i(t+\tau/2,\vec{x}+\vec{y}/2)\cdot {\sf H}^2_{jm}(t-\tau/2,\vec{x}-\vec{y}/2)\psi^{*}_m(t-\tau/2,\vec{x}-\vec{y}/2)]+\ldots\,,
\end{align}
the right-hand side of equation \eqref{eq:qktimedeppotstep1} also contains terms proportional to the time-derivative of the Hamiltonian,
\begin{align}
\label{eq:qktimedeppotstep2} 
 \ldots &+\frac12\int d\tau e^{i\epsilon\tau} \int d\vec{y} e^{-i\vec{p}\vec{y}}  \nonumber\\ 
&\times [\partial_t{\sf H}_{in}(t+\tau/2,\vec{x}+\vec{y}/2) \psi_n(t+\tau/2,\vec{x}+\vec{y}/2)\cdot \psi^{*}_j(t-\tau/2,\vec{x}-\vec{y}/2)
\nonumber\\
&+\psi_i(t+\tau/2,\vec{x}+\vec{y}/2)\cdot \partial_t{\sf H}_{jm}(t-\tau/2,\vec{x}-\vec{y}/2)\psi^{*}_m(t-\tau/2,\vec{x}-\vec{y}/2)]\,.
\end{align}

Written in this form, the evolution equation for the two-point correlator is valid for generic potentials. 
For a time-dependent but 
homogeneous potential the  kinetic and potential energy operators commute and the action of the Hamiltonian amounts, in the 
momentum representation, to multiplication  of the wave function by ${\sf H}_{ij}(t,\vec{p})=\delta_{ij}\,\omega_i(\vec{p})+{\sf V}_{ij}(t)$. 
Expressing  the coordinate-representation wave functions in equations \eqref{eq:qktimedeppotstep1} and 
\eqref{eq:qktimedeppotstep2}  in terms of their momentum-representation counterparts and subsequently introducing the 
center and relative momenta we obtain 
\begin{align}
\label{eq:qktimedeppotstep3} 
\epsilon\,\partial_t \varrho_{ij}&(t,\vec{x},\epsilon,\vec{p})  =
-\frac{i}2\int d\tau e^{i\epsilon\tau} \int d\vec{y} e^{-i\vec{p}\vec{y}}  
\int \frac{d^3\vec{s}}{(2\pi)^3} \frac{d^3\vec{\Delta}}{(2\pi)^3}e^{i\vec{\Delta}\vec{x}}e^{i\vec{s}\vec{y}}\nonumber\\ 
&\times [{\sf H}^2_{in}(t+\tau/2,\vec{s}+\vec{\Delta}/2) \psi_n(t+\tau/2,\vec{s}+\vec{\Delta}/2)
\psi^{*}_j(t-\tau/2,\vec{s}-\vec{\Delta}/2)\nonumber\\
&-\psi_i(t+\tau/2,\vec{s}+\vec{\Delta}/2) \psi^{*}_m(t-\tau/2,\vec{s}-\vec{\Delta}/2){\sf H}^2_{mj}(t-\tau/2,\vec{s}-\vec{\Delta}/2)]
+\ldots
\end{align}
and 
\begin{align}
\label{eq:qktimedeppotstep4} 
\ldots &+\frac12\int d\tau e^{i\epsilon\tau} \int d\vec{y} e^{-i\vec{p}\vec{y}}  
\int \frac{d^3\vec{s}}{(2\pi)^3} \frac{d^3\vec{\Delta}}{(2\pi)^3}e^{i\vec{\Delta}\vec{x}}e^{i\vec{s}\vec{y}}\nonumber\\ 
&\times [\partial_t{\sf V}_{in}(t+\tau/2) \psi_n(t+\tau/2,\vec{s}+\vec{\Delta}/2)
\psi^{*}_j(t-\tau/2,\vec{s}-\vec{\Delta}/2)
\nonumber\\
&+\psi_i(t+\tau/2,\vec{s}+\vec{\Delta}/2) \psi^{*}_m(t-\tau/2,\vec{s}-\vec{\Delta}/2)
\partial_t{\sf V}_{mj}(t-\tau/2)]	
\end{align}
respectively. Proceeding as above we expand the Hamiltonian to the first order in the gradients in 
equation \eqref{eq:qktimedeppotstep3}, ${\sf H}^2_{ij}(t\pm\tau/2,\vec{s}\pm\vec{\Delta}/2) \approx {\sf H}^2_{ij}(t,\vec{s}) \pm
\partial_\vec{s}{\sf H}^2_{ij}(t,\vec{s})\vec{\Delta}/2\pm\partial_t {\sf H}^2_{ij}(t,\vec{s})\tau/2$. Subsequently using 
$e^{i\vec{\Delta}\vec{x}}\vec{\Delta}=-i\partial_\vec{x} e^{i\vec{\Delta}\vec{x}}$ and $e^{i\epsilon\tau}\tau = 
-i\partial_\epsilon e^{i\epsilon\tau}$ we obtain from equation \eqref{eq:qktimedeppotstep3} 
\begin{align}
\label{eq:qktimedeppotstep5} 
\epsilon\partial_t\varrho(t,\vec{x},\epsilon,\vec{p})  \ni & 
-{\textstyle\frac14}\{\partial_\vec{p}{\sf H}^2(t,\vec{p}),\partial_\vec{x}\varrho(t,\vec{x},\epsilon,\vec{p})\}
-{\textstyle\frac14}\{\partial_t{\sf H}^2(t,\vec{p}),\partial_\epsilon\varrho(t,\vec{x},\epsilon,\vec{p})\}\nonumber\\
&-{\textstyle\frac{i}{2}}[{\sf H}^2(t,\vec{p}),\varrho(t,\vec{x},\epsilon,\vec{p})]\,.
\end{align}
Because equation \eqref{eq:qktimedeppotstep4}  already contains a derivative of the potential, to the first order in the 
gradient expansion it is sufficient to 
approximate 	$\partial_t{\sf V}_{in}(t\pm\tau/2) $ by $\partial_t{\sf V}_{in}(t)$. In this approximation we obtain from 
equation \eqref{eq:qktimedeppotstep4} 
\begin{align}
\label{eq:qktimedeppotstep6} 
\epsilon\partial_t\varrho(t,\vec{x},\epsilon,\vec{p})  \ni  {\textstyle\frac12}\{\partial_t{\sf V}(t),\varrho(t,\vec{x},\epsilon,\vec{x})\}\,.
\end{align}
Combining equations \eqref{eq:qktimedeppotstep5}  and \eqref{eq:qktimedeppotstep6} we finally arrive at  
\begin{align}
\label{eq:qktimedeppot} 
\epsilon\partial_t\varrho(t,\vec{x},\epsilon,\vec{p})  & 
+{\textstyle\frac14}\{\partial_\vec{p}{\sf H}^2(t,\vec{p}),\partial_\vec{x}\varrho(t,\vec{x},\epsilon,\vec{p})\}
+{\textstyle\frac14}\{\partial_t{\sf H}^2(t,\vec{p}),\partial_\epsilon\varrho(t,\vec{x},\epsilon,\vec{p})\}\nonumber\\
& - {\textstyle\frac12}\{\partial_t{\sf V}(t),\varrho(t,\vec{x},\epsilon,\vec{x})\}
\approx -{\textstyle\frac{i}{2}}[{\sf H}^2(t,\vec{p}),\varrho(t,\vec{x},\epsilon,\vec{p})]\,.
\end{align}
In appendix \ref{sec:timedeppot} we rederive the left-hand side of equation \eqref{eq:qktimedeppot} for a single 
neutrino generation using the gradient expansion to obtain an explicit expression for the two-point correlator.
This expression illu\-strates that a homogeneous but time-dependent potential affects 
primarily the spectrum of the two-point correlator. Furthermore, the terms 
$\{\partial_t{\sf H}^2(t,\vec{p}),\partial_\epsilon\varrho(t,\vec{x},\epsilon,\vec{p})\}$ and 
$\{\partial_t{\sf V}(t),\varrho(t,\vec{x},\epsilon,\vec{x})\}$ have a common origin and stem from time-derivative of 
the energy-conserving delta-function. Therefore, even though at first sight the $\{\partial_t{\sf V}(t),\varrho(t,\vec{x},\epsilon,\vec{x})\}$
term seems to describe effects of absorption or loss of coherence, it actually describes shifts of the spectrum induced 
by the time-dependent potential, just like the term $\{\partial_t{\sf H}^2(t,\vec{p}),\partial_\epsilon\varrho(t,\vec{x},\epsilon,\vec{p})\}$.

\paragraph{Recovering relativistic limit.}
Since we work to linear order in the gradient expansion here, the qu\-antum kinetic equation for neutrinos propagating in a general 
potential can be deduced by combining equations \eqref{eq:qkeqinhompot} and \eqref{eq:qktimedeppot},
\begin{align}
\label{eq:qkeqgenericpot}
\epsilon\,\partial_t \varrho(t,\vec{x},\epsilon,\vec{p}) & +
{\textstyle\frac14}\{\partial_\vec{p}{\sf H}^2(t,\vec{x},\vec{p}),\partial_\vec{x}\varrho(t,\vec{x},\epsilon,\vec{p})\}
-{\textstyle\frac14}\{\partial_\vec{x}{\sf H}^2(t,\vec{x},\vec{p}),\partial_\vec{p}\varrho(t,\vec{x},\epsilon,\vec{p})\}\nonumber\\
&+{\textstyle\frac14}\{\partial_t{\sf H}^2(t,\vec{x},\vec{p}),\partial_\epsilon\varrho(t,\vec{x},\epsilon,\vec{p})\} 
- {\textstyle\frac12}\{\partial_t{\sf V}(t,\vec{x}),\varrho(t,\vec{x},\epsilon,\vec{x})\} \nonumber\\
&\approx -{\textstyle\frac{i}{2}}[{\sf H}^2(t,\vec{x},\vec{p})+{\textstyle \frac{i}2}[\partial_\vec{x}{\sf V}(t,\vec{x}),\partial_\vec{p}
\omega(\vec{p})],\varrho(t,\vec{x},\epsilon,\vec{p})]\,.
\end{align}
To conclude our analysis of the evolution equation for the two-point correlator we demonstrate how to 
recover equation \eqref{eq:ultrarelapprox}, derived in the relativistic limit, from equation \eqref{eq:qkeqgenericpot}, which 
is valid for any value of the momentum.

To handle the first anticommutator on the left-hand side of equation \eqref{eq:qkeqgenericpot} we use the identity 
$\{\partial_\vec{p} {\sf H}^2,\partial_\vec{x} \varrho\}=\{{\sf H},\{\partial_\vec{p}{\sf H},\partial_\vec{x}\varrho\}\}
+[\partial_\vec{p}{\sf H},[{\sf H},\partial_\vec{x}\varrho]]$. The term  $\{{\sf H},\{\partial_\vec{p}{\sf H},\partial_\vec{x}\varrho\}\}$
can be approximated by $\{{\sf \omega},\{\partial_\vec{p}{\sf H},\partial_\vec{x}\varrho\}\}_{ij}=(\omega_i+\omega_j)
\{\partial_\vec{p}{\sf H},\partial_\vec{x}\varrho\}_{ij}$. Similarly, the  term $[\partial_\vec{p}{\sf H},[{\sf H},\partial_\vec{x}\varrho]]$
can be approxi\-mated by $[\partial_\vec{p}\omega,[\omega,\partial_\vec{x}\varrho]]_{ij}=\Delta\vec{v}_{ij}\Delta\omega_{ij}
\partial_\vec{x}\varrho$. In the relativistic limit the second term is negligibly small compared to the first.

For the second anticommutator we use $\{\partial_\vec{x} {\sf H}^2,\partial_\vec{p} \varrho\}=
\{{\sf H},\{\partial_\vec{x}{\sf H},\partial_\vec{p}\varrho\}\}+[\partial_\vec{x}{\sf H},[{\sf H},\partial_\vec{p}\varrho]]$.
In the first term we again use the approximation $\{{\sf \omega},\{\partial_\vec{x}{\sf H},\partial_\vec{p}\varrho\}\}_{ij}=
(\omega_i+\omega_j)\{\partial_\vec{x}{\sf H},\partial_\vec{p}\varrho\}_{ij}$. To the first order in the potential the second
term can be approximated by $[\partial_\vec{x}{\sf V},[{\sf \omega},\partial_\vec{p}\varrho]]$ which vanishes in the limit 
of  $\omega\propto {\sf 1}$. Hence this term is proportional to $\Delta\omega$ and is again negligibly small compared 
to the first one. Analogously $\{\partial_t {\sf H}^2,\partial_\epsilon \varrho\}=
\{{\sf H},\{\partial_t{\sf H},\partial_\epsilon\varrho\}\}+[\partial_t{\sf H},[{\sf H},\partial_\epsilon\varrho]]$. The first term 
can be approximated by  $\{{\sf \omega},\{\partial_t{\sf H},\partial_\epsilon\varrho\}\}_{ij}=
(\omega_i+\omega_j)\{\partial_t{\sf H},\partial_\epsilon\varrho\}_{ij}$, and the second, to the first order in the potential,
by $[\partial_t{\sf V},[{\sf \omega},\partial_\vec{p}\varrho]]$. The second term is proportional to $\Delta\omega$ and, 
in the relativistic limit, negligibly small compared to the first.

On the right-hand side of equation \eqref{eq:qkeqgenericpot} we use the identity $[{\sf H}^2,\varrho]\!=\!\{{\sf H},[{\sf H},\varrho]\}$
to handle the commutator, and further approximate it by $\{{\sf \omega},[{\sf H},\varrho]\}_{ij}=(\omega_i+\omega_j)[{\sf H},\varrho]_{ij}$.

Common to all the (leading) terms discussed so far is the overall factor $\omega_i+\omega_j$. The two new 
terms, $\{\partial_t{\sf V}(t,\vec{x}),\varrho(t,\vec{x},\epsilon,\vec{x})\}$ one the left-hand side and 
$[\partial_\vec{x}{\sf V}(t,\vec{x}),\partial_\vec{p}\omega(\vec{p})]$ on the right-hand side, do not share this feature.
Therefore, they can be neglected in the relativistic limit.

As has been discussed in section \ref{sec:liouvilleterm}, barring small off-shell contributions $2\epsilon\approx 
\omega_i+\omega_j$. Dividing the left- and right-hand side of equation \eqref{eq:qkeqgenericpot} by $\omega_i+\omega_j$
we therefore recover equation \eqref{eq:ultrarelapprox}. It is also worth mentioning that the subleading contribution  
$\Delta\vec{v}_{ij}(\omega_i-\omega_j)/(\omega_i+\omega_j)/2$ that stems from the anticommutator
$\{\partial_\vec{p}{\sf H}^2(t,\vec{x},\vec{p}),\partial_\vec{x}\varrho(t,\vec{x},\epsilon,\vec{p})\}$ is equal to 
$\vec{v}_{ij}-\vec{\bar v}_{ij}$. This explains why the center of  momentum velocity is replaced by the average velocity in 
the relativistic approximation.

\paragraph{Quantum-kinetic equation.}
Depending on its nature the neutrino possesses either two (Majorana particle) or four (Dirac particle) 
components. On the other hand, the wave function in the quantum-mechanical approach to neutrino oscillation describes 
a single component, either a left-handed neutrino or a right-handed antineutrino. For typical experimental setups this is 
well motivated because the spin-flipping (or active-sterile) conversions are strongly suppressed by the smallness of the 
$m/\omega$ ratio. Nevertheless, strictly speaking, this means that the quantum-field theoretical counterpart of the conventional
quantum-mechanical approach to neutrino oscillations is the theory of mixing scalar fields. To the first order in the gradient 
expansion the quantum kinetic equation for the respective Wightman propagator reads
\begin{align}
\label{eq:qkeqwignspacefinal}
\epsilon\partial_t \varrho_<(t,\vec{x},\epsilon,\vec{p}) & 	+ 
\textstyle{\frac14}\{\partial_\vec{p}{\sf H}^2(t,\vec{x},\vec{p}),\partial_\vec{x}\varrho_<(t,\vec{x},\epsilon,\vec{p})\} - 
\textstyle{\frac14}\{\partial_\vec{x}{\sf H}^2(t,\vec{x},\vec{p}),\partial_\vec{p}\varrho_<(t,\vec{x},\epsilon,\vec{p})\} \nonumber\\
& +\textstyle{\frac14}\{\partial_t{\sf H}^2(t,\vec{x},\vec{p}),\partial_\epsilon \varrho_<(t,\vec{x},\epsilon,\vec{p})\}
\approx - {\textstyle\frac{i}2}[{\sf H}^2(t,\vec{x},\vec{p}),\varrho_<(t,\vec{x},\epsilon,\vec{p})]\,,
\end{align}
with the effective Hamiltonian defined as 
\begin{align}
{\sf H}^2(t,\vec{x},\vec{p}) \equiv \vec{p}^2 + m^2 + \Sigma(t,\vec{x}) = \omega^2(\vec{p}) + \Sigma(t,\vec{x})\,,
\end{align}
see appendix \ref{sec:quantumkineticeq} for more details.
The local self energy $\Sigma(t,\vec{x})$ models the lowest-order contribution to the neutrino self-energy evaluated in the limit 
$m_{W,Z}\rightarrow \infty$ \cite{Vlasenko:2013fja}.

The form of equation \eqref{eq:qkeqwignspacefinal} is  almost identical to the form of equation \eqref{eq:qkeqgenericpot}, 
except for the absence of the two new terms, $\{\partial_t{\sf V}(t,\vec{x}),\varrho(t,\vec{x},\epsilon,\vec{x})\}$ and 
$[[\partial_\vec{x}{\sf V}(t,\vec{x}),\partial_\vec{p}\omega(\vec{p})],\varrho(t,\vec{x},\epsilon,\vec{p})]$. Their absence can be traced back to the slightly different 
gradient expansion schemes used to derive equations \eqref{eq:qkeqgenericpot} and \eqref{eq:qkeqwignspacefinal}.
Whereas to derive the former we expanded the potential ${\sf V}$, to derive the latter we expanded the self-energy 
$\Sigma$. The self-energy can be approximately expressed through ${\sf V}$ as $\Sigma \sim \omega V + V \omega$. 
Therefore, appearance of terms containing a different combination of $\omega$ and ${\sf V}$ in the quantum kinetic equation 
is not possible.  As has been argued above, in the relativistic limit the two new terms are subdominant 
and can be safely neglected anyway.

\subsection{\label{sec:kineticsmed}Description in terms of the Wigner function}
The (relativistic) evolution equation for the two-point correlator \eqref{eq:ultrarelapprox} can be trivially 
integrated over $\epsilon$. The resulting evolution equation for the Wigner function  matches (the collisionless limit of) 
the kinetic equation of Sigl and Raffelt and admits solutions consistent with the uncertainty principle also for neutrinos 
propagating in matter.

\paragraph{Kinetic equation.}
To derive evolution equation for the Wigner function we integrate equation \eqref{eq:ultrarelapprox} 
over $\epsilon$. Because 
$\varrho(t,\vec{x},\epsilon,\vec{p})$ vanishes for $\epsilon\rightarrow\pm\infty$, the last term on the left-hand side vanishes 
upon the integration. Integration of the remaining terms yields 
\begin{align}
\label{eq:siglraffelt}
\partial_t \varrho(t,\vec{x},\vec{p})  
&+{\textstyle\frac12}\{\partial_\vec{p} {\sf H}(t,\vec{x},\vec{p}),\partial_\vec{x}\varrho(t,\vec{x},\vec{p})\}
-{\textstyle\frac12}\{\partial_\vec{x} {\sf H}(t,\vec{x},\vec{p}),\partial_\vec{p}\varrho(t,\vec{x},\vec{p})\}\nonumber\\ 
&\approx -i\,[{\sf H}(t,\vec{x},\vec{p}),\varrho(t,\vec{x},\vec{p})]\,,
\end{align}
valid in the physically relevant relativistic limit. The form of equation \eqref{eq:siglraffelt} matches the (collisionless limit of the)
kinetic equation of Sigl and Raffelt \cite{Sigl:1992fn}.  As has been emphasized in reference \cite{Stirner:2018ojk}, in the
relativistic limit the first anticommutator on the left-hand side of equation \eqref{eq:siglraffelt} contains the full flavor-dependent 
velocity structure of the Liouville term.

\paragraph{Neutrino propagation in a constant potential.} 
For a constant potential the second anticommutator on the right-hand side of equation \eqref{eq:siglraffelt} vanishes and 
it simplifies to 
\begin{align}
\label{eq:siglraffeltconstpot}
\partial_t \varrho(t,\vec{x},\vec{p})  
&+{\textstyle\frac12}\{\partial_\vec{p} {\sf H}(\vec{p}),\partial_\vec{x}\varrho(t,\vec{x},\vec{p})\}
\approx -i\,[{\sf H}(\vec{p}),\varrho(t,\vec{x},\vec{p})]\,.
\end{align}

We can construct its solution by substituting the (well known) solution of the 
Schr\"odinger equation in a constant potential, $\psi(t,\vec{p}) = e^{-i{\sf H}(\vec{p})t} \psi(0,\vec{p})$, into equation 
\eqref{eq:wignerfunction}. This results in
\begin{align}
\label{eq:wignerfuncconstpot}
\varrho(t,\vec{x},\vec{p})=\int\frac{d^3\vec{\Delta}}{(2\pi)^3}e^{i\vec{\Delta}\vec{x}}
e^{-i{\sf H}(\vec{p}+\vec{\Delta}/2)t}\psi(0,\vec{p}+{\textstyle\frac12}\vec{\Delta})\psi^\dagger(0,\vec{p}-{\textstyle\frac12}\vec{\Delta})
e^{i{\sf H}(\vec{p}-\vec{\Delta}/2)t}\,.
\end{align}
Equation \eqref{eq:wignerfuncconstpot} only approximately solves equation \eqref{eq:siglraffeltconstpot}. Differentiating 
the former with respect to time we obtain $-i{\sf H}(\vec{p}+\vec{\Delta}/2)e^{-i{\sf H}(\vec{p}+\vec{\Delta}/2)t}$ and 
$e^{i{\sf H}(\vec{p}-\vec{\Delta}/2)t}i{\sf H}(\vec{p}-\vec{\Delta}/2)$ under the integral. To recover equation 
\eqref{eq:siglraffeltconstpot} we further need to perform a gradient expansion of the pre-exponential factors,
${\sf H}(\vec{p}\pm\vec{\Delta}/2)\approx {\sf H}(\vec{p})\pm\partial_\vec{p}{\sf H}(\vec{p})\vec{\Delta}/2$ (and 
afterwards use $\vec{\Delta}e^{i\vec{\Delta}\vec{x}}=-i\partial_{\vec{x}}e^{i\vec{\Delta}\vec{x}}$). On the other 
hand, if the gradient expansion of the Hamiltonian is performed directly in the integrand of equation 
\eqref{eq:wignerfuncconstpot} then the resulting expression solves the evolution equation. In other words, the gradient 
expansion approximation used to obtain   equation \eqref{eq:siglraffeltconstpot} propagates also to its 
solutions. For a Hamiltonian of the form ${\sf H}(\vec{p})=\omega(\vec{p})+{\sf V}$ this implies that the velocity 
$\partial_\vec{p} {\sf H}(\vec{p})$ matches the vacuum propagation velo\-ci\-ty $\partial_\vec{p} \omega(\vec{p})$. 
Hence, equation \eqref{eq:siglraffeltconstpot} does not account for matter corrections to the 
neutrino propagation velocity. For relativistic neutrinos this approximation is nevertheless more than adequate.

By analogy with equation \eqref{eq:shapepahsewigner} we tentatively write the Wigner function in the form  
\begin{align}
\label{eq:wignerfuncconstpotappr}
\varrho(t,\vec{x},\vec{p}) = e^{-i{\sf H}(\vec{p})t} g(t,\vec{x},\vec{p}) e^{i{\sf H}(\vec{p})t}\,.
\end{align}
From equations \eqref{eq:siglraffeltconstpot} and \eqref{eq:wignerfuncconstpotappr} it then follows that the shape factor 
satisfies
\begin{align}
\label{eq:shapeeqconstpot} 
\partial_t g(t,\vec{x},\vec{p}) + {\textstyle \frac12} \{{\bm \upsilon}(t,\vec{p}),\partial_{\vec{x}}g(t,\vec{x},\vec{p})\} = 0\,,
\end{align}
with the effective velocity ${\bm \upsilon}(t,\vec{p})$ given by 
\begin{align}
\label{eq:effectivevelocity}
{\bm \upsilon}(t,\vec{p}) = e^{i{\sf H}(\vec{p})t}\partial_\vec{p}{\sf H}(\vec{p})e^{-i{\sf H}(\vec{p})t}\,.
\end{align}
Performing a gradient expansion of the Hamiltonian in equation \eqref{eq:wignerfuncconstpot} and subsequently using the
Feynman disentanglement theorem we obtain for the shape factor
\begin{align}
\label{eq:shapefactconstpotappr}
g(t,\vec{x},\vec{p}) & = \int\frac{d^3\vec{\Delta}}{(2\pi)^3} e^{i\vec{\Delta}\vec{x}}\,
T_c\,e^{-i\int_0^t{\bm \upsilon}(t,\vec{p})\vec{\Delta}/2\,dt}\psi(0,\vec{p}+\textstyle{\frac12}\vec{\Delta})\nonumber\\
&\times \psi^\dagger(0,\vec{p}-\textstyle{\frac12}\vec{\Delta})\,
T_a\,e^{-i\int^t_0{\bm \upsilon}(t,\vec{p})\vec{\Delta}/2\,dt}\,,
\end{align}
with $T_c$ and $T_a$ being the time-ordering and the anti-time-ordering operators respectively. As can be verified by 
substitution, equation \eqref{eq:shapefactconstpotappr} is consistent with equations \eqref{eq:shapeeqconstpot} and 
\eqref{eq:effectivevelocity}.

\paragraph{The uncertainty principle.} 
The matrices ${\sf H}(\vec{p})$ and $\partial_\vec{p}{\sf H}(\vec{p})$ do not commute in general. The lowest order 
approximation to the effective velocity consists in neglecting $[{\sf H}(\vec{p}),\partial_\vec{p}{\sf H}(\vec{p})]$ in 
equation \eqref{eq:effectivevelocity}. Parametrically this commutator is of the order of $\sin{2\theta}{\sf V}\Delta \vec{v}_{ij}$, 
where $\theta$ is the vacuum mixing angle, and is therefore strongly suppressed with respect to the leading term 
$\partial_\vec{p}{\sf H}(\vec{p})$. It becomes comparable to $\partial_\vec{p}{\sf H}(\vec{p})$ only at late times, 
$\sin{2\theta}{\sf V} |\Delta \vec{v}_{ij}| t \sim 1$. For ${\sf V}\sim 1$ eV, $\omega\sim 1$~MeV, and the solar mass 
difference this amounts to a distance that is an order of magnitude larger than the solar radius. This approximation results 
in ${\bm \upsilon}(t,\vec{p}) \approx \partial_\vec{p}{\sf H}(\vec{p})=\partial_{\vec{p}} \omega(\vec{p})=\vec{v}(\vec{p})$ 
and shape factor of the form 
\begin{align}
\label{eq:shapefactconstpotlowestord}
g(t,\vec{x},\vec{p}) & \approx \int\frac{d^3\vec{\Delta}}{(2\pi)^3} e^{i\vec{\Delta}\vec{x}}
e^{-i\vec{v}(\vec{p})\vec{\Delta}t/2}\psi(0,\vec{p}+\textstyle{\frac12}\vec{\Delta}) \psi^\dagger(0,\vec{p}-\textstyle{\frac12}\vec{\Delta})
e^{-i\vec{v}(\vec{p})\vec{\Delta}t/2}\,.
\end{align}
For Gaussian initial conditions the approximate shape factor equation \eqref{eq:shapefactconstpotlowestord} takes 
the form identical to equation \eqref{eq:shapewignergaussian}, up to the (negligible in the relativistic limit) 
replacement $\vec{v}_{ij} \rightarrow \bar{\vec{v}}_{ij}$, 
\begin{align}
\label{eq:shapfactconstpot}
g_{ij}&(t,\vec{x},\vec{p}) \propto \exp\biggl(-\frac{(\vec{p}-\vec{p}_w)^2}{2\sigma_p^2}\biggr)\,
\exp\biggl(-\frac{(\vec{\bar{v}}_{ij}(\vec{p})t-\vec{x})^2}{2\sigma^2_x}\biggr)\,.
\end{align}
As has been argued in section \ref{sec:kinetics}, this shape factor is consistent with the uncertainty principle. The same 
applies to Lorentzian initial conditions, see appendix \ref{sec:lorentzian}. In other words, also for neutrino propagating in 
matter the evolution equation admits solutions consistent with the uncertainty principle.

\paragraph{Kinematical decoherence.}
As follows from equations \eqref{eq:wignerfuncconstpotappr} and \eqref{eq:shapfactconstpot}, also in matter
each individual momentum mode of the Wigner function does not experience any suppression in the course of
collisionless neutrino
propagation. On the other hand, their growing dephasing results in kinematical decoherence in the density matrix at late times. 
To demonstrate this we derive the latter by integrating equation \eqref{eq:wignerfuncconstpotappr} over $\vec{p}$.  
For Gaussian initial conditions the integral can be estimated using the saddle point approximation. To this end we use 
$e^{\mp i{\sf H}(\vec{p})t}={\sf U}(\vec{p}) e^{\mp i{\sf E}(\vec{p})t} {\sf U}^\dagger(\vec{p})$, where ${\sf U}(\vec{p})$ is a
unitary matrix diagonalizing the Hamiltonian, ${\sf U}^\dagger(\vec{p}) {\sf H}(\vec{p}){\sf U}(\vec{p})={\sf E}(\vec{p})$. 
Expanding ${\sf E}(\vec{p})$ in the vicinity of $\vec{p}_w$, ${\sf E}(\vec{p}) \approx {\sf E}(\vec{p}_w)+\vec{u}(\vec{p}_w)
(\vec{p}-\vec{p}_w)$, where $\vec{u}(\vec{p})\equiv \partial_{\vec{p}}{\sf E}(\vec{p})$ denotes group velocity of the 
propagation eigenstates, and taking into account that ${\sf E}(\vec{p})$ and $\vec{u}(\vec{p})$ commute, we find
for the density matrix
\begin{align}
\label{eq:densmatrconstpot} 
\rho(t,\vec{x}) \approx e^{- i{\sf H}(\vec{p}_w)t} g(t,\vec{x}) e^{ i{\sf H}(\vec{p}_w)t}\,,
\end{align}
with the shape factor given (for the Gaussian initial conditions) by
\begin{align}
\label{eq:shapefactorconstpot}
g_{ij}(t,\vec{x}) \approx {\sf U}_{im}(\vec{p}_w){\sf U}^\dagger_{mk}(\vec{p}_w)
g_{kl}(t,\vec{x},\vec{p}_w)  {\sf U}_{ln}(\vec{p}_w){\sf U}^\dagger_{nj}(\vec{p}_w)
\exp\left(-\frac{\Delta \vec{u}^2_{mn}(\vec{p}_w) t^2}{8\sigma^2_x}\right)\,.
\end{align}  
Hence, for $\Delta \vec{u} \neq 0	$ oscillations are  suppressed at late times. 
In vacuum equation \eqref{eq:densmatrconstpot} reverts to equation \eqref{eq:densmatrwignerapprox}. 

Let us emphasize that even though the evolution equation \eqref{eq:siglraffeltconstpot} neglects medium corrections to 
the neutrino propagation velocity, it does take into account medium corrections to the group velocity of the propagation eigenstates.  
Close to the Mikheyev--Smirnov--Wolfenstein  resonance \cite{Mikheyev:1989dy} the latter, though small in the absolute value, 
are of the same order of magnitude as the differe\-nce of the vacuum neutrino velocities.  As a result $\Delta\vec{u}_{mn}$ can 
vanish. Therefore, matter effects can suppress development of the \emph{kinematical} decoherence in the density matrix 
under certain circumstances, see e.g.	\cite{Mikheyev:1989dy,Kersten:2015kio} and references therein. One should however
keep in mind, that scattering on the matter particles induces \emph{dynamical} decoherence in the Wigner function and density 
matrix \cite{Richers:2019grc}.
	
\paragraph{Neutrino propagation in a time-dependent potential.} As the width of the neutrino wave packet in coordinate 
space is typically much smaller than the length scale for the variation of the matter potential in the interior of the Sun or the Earth, 
one can approximate neutrino propagation by propagation in a spatially homogeneous but time-dependent potential, which is 	
governed by 
\begin{align}
\label{eq:siglraffelttimedeppot}
\partial_t \varrho(t,\vec{x},\vec{p})  
&+{\textstyle\frac12}\{\partial_\vec{p} {\sf H}(t,\vec{p}),\partial_\vec{x}\varrho(t,\vec{x},\vec{p})\}
\approx -i\,[{\sf H}(t,\vec{p}),\varrho(t,\vec{x},\vec{p})]\,.
\end{align}
The value of this effective matter potential at a time $t$ is given by the potential at the position the neutri\-no wave packet reaches by 
that time. This approximation is widely used in the literature, see e.g. references \cite{Maltoni:2015kca,Akhmedov:2017mcc}.

Because the Hamiltonian at different times does not in general commute with itself, the solution of the Schr\"odinger equation 
contains the time-ordered exponent, \smash{$\psi(t,\vec{p})=T_c\,e^{-i\int^t_0 {\sf H}(t,\,\vec{p})dt} \psi(0,\vec{p})$}. 
Therefore, the evolution operators $e^{-i{\sf H}(\vec{p}+\vec{\Delta}/2)t}$ and $e^{i{\sf H}(\vec{p}-\vec{\Delta}/2)t}$ in 
equation \eqref{eq:wignerfuncconstpot} are replaced by $T_c\,e^{-i\int^t_0{\sf H}(t,\,\vec{p}+\vec{\Delta}/2)dt}$ and 
$T_a\,e^{i\int^t_0{\sf H}(t,\,\vec{p}-\vec{\Delta}/2)dt}$,
\begin{align}
\label{eq:wignerfunctimedeptpot}
\varrho(t,\vec{x},\vec{p})&=\!\int\!\frac{d^3\vec{\Delta}}{(2\pi)^3}e^{i\vec{\Delta}\vec{x}}\,
 T_c\, e^{-i\int^t_0{\sf H}(t,\,\vec{p}+\vec{\Delta}/2) dt} \psi(0,\vec{p}+{\textstyle \frac12}\vec{\Delta}) \nonumber\\
& \times \psi^\dagger(0,\vec{p}-{\textstyle\frac12}\vec{\Delta})
T_a\, e^{i\int^t_0{\sf H}(t,\,\vec{p}-\vec{\Delta}/2) dt}\,.
\end{align} 
Equation \eqref{eq:wignerfunctimedeptpot} only approximately solves equation \eqref{eq:siglraffelttimedeppot}.
Differentiating the former with respect to $t$ we obtain 
\smash{$-i{\sf H}(t,\,\vec{p}+\vec{\Delta}/2) T_c\, e^{-i\int^t_0{\sf H}(t,\,\vec{p}+\vec{\Delta}/2) dt}$}
and \smash{$T_a\, e^{i\int^t_0{\sf H}(t,\,\vec{p}-\vec{\Delta}/2) dt}\,i{\sf H}(t,\,\vec{p}-\vec{\Delta}/2)$} under the 
integral sign. Similar to the case of a constant potential, to recover equation \eqref{eq:siglraffelttimedeppot}
we  need to perform a gradient expansion of the pre-exponential factors, 
${\sf H}(t,\vec{p}\pm\vec{\Delta}/2)\approx {\sf H}(t,\vec{p})\pm\partial_\vec{p}{\sf H}(t,\vec{p})\vec{\Delta}/2$ (and 
subsequently use $\vec{\Delta}e^{i\vec{\Delta}\vec{x}}=-i\partial_{\vec{x}}e^{i\vec{\Delta}\vec{x}}$). This implies that  
the solution of equation \eqref{eq:siglraffelttimedeppot} is obtained from equation \eqref{eq:wignerfunctimedeptpot} 
by performing a gradient expansion of the Hamiltonian in the (anti-) time-ordered exponents.

Proceeding similarly to the case of a constant potential we tentatively write the Wigner function in the form 
\begin{align}
\label{eq:wignerfunctimedeppotappr}
\varrho(t,\vec{x},\vec{p}) \approx T_c\, e^{-i\int^t_0 {\sf H}(t,\vec{p})dt} g(t,\vec{x},\vec{p}) \,
T_a\,e^{i\int^t_0{\sf H}(t,\vec{p})dt}\,.
\end{align}
From equations \eqref{eq:siglraffelttimedeppot} and \eqref{eq:wignerfunctimedeppotappr} it follows that the 
shape factor satisfies equation \eqref{eq:shapeeqconstpot}  with the effective velocity now given by 
\begin{align}
\label{eq:effectveltimedeppot}
{\bm \upsilon}(t,\vec{p}) = T_a\, e^{i\int^t_0{\sf H}(t,\vec{p})dt}\,\partial_\vec{p}{\sf H}(t,\vec{p})\,
T_c\,e^{-i\int^t_0{\sf H}(t,\vec{p})dt}\,.
\end{align}
Performing a gradient expansion of the Hamiltonian in equation \eqref{eq:wignerfunctimedeptpot} and further using the
Feynman disentanglement theorem we obtain for the shape factor an expression identical to equation 
\eqref{eq:shapefactconstpotappr} but with the effective velocity given by equation \eqref{eq:effectveltimedeppot}.
 
\section{\label{sec:conclusions}Summary and conclusion}
The present work establishes a connection between three approaches to the description of neutrino oscillations:  
(i) a quantum-mechanical approach operating with the neutrino wave function or density matrix, (ii) a quantum kinetic 
approach operating with the Wightman propagator, and (iii) a kinetic approach operating with the matrix of densities.

The quantum-mechanical approach is well suited to analysis of neutrino oscillations in vacuum and its collisionless  
propagation in matter. In particular, it is automatically consistent with the uncertainty principle and, as a result, takes into 
account the effect of wave packet separation. On the other hand, it is not capable of accounting for scattering processes. 
This limits its applicability to analysis of neutrino propagation in exploding supernovae, where oscillations and collisions may 
be equally important. 

The quantum kinetic equation for the Wightmann propagator is derived using methods of non-equilibrium quantum field theory. 
An evolution equation for its quantum-mechanical counterpart -- the two-point correlator -- can be derived from the 
Schr\"odinger equation. For neutrino propagation in vacuum these two (quite distinct) approaches lead to identical equations. 
For collisionless neutrino propagation in matter the resulting equations differ only by subleading terms which can be traced back 
to slightly different gradient expansion schemes. Analysis of the two-point correlator for typical initial conditions reveals that in 
addition to the usual mass shells $\epsilon = \omega_i$ there are  $\epsilon = (\omega_i + \omega_j)/2$ shells responsible for 
neutrino oscillations. Their existence explains why the Liouville term of the evolution (as well as of the quantum kinetic) equation, 
$\partial_t + \vec{v}\partial_\vec{x}$, does not carry any generation indices. The propagation velocity $\vec{v}=\vec{p}/\epsilon$ 
is determined by the spectrum of the two-point correlator. In the propagation basis its diagonal elements peak  in the vicinity of 
the respective mass shells and propagate with velocity $\vec{v}\approx \vec{p}/\omega_i$. The off diagonal elements peak in 
the vicinity of the oscillation shells and propagate with the center of momentum velocity 
$\vec{v} \approx 2\vec{p}/(\omega_i+\omega_j)$.
 
For typical initial conditions the off-shell contributions are negligible and it is sufficient to keep track of the evolution of the 
two-point correlator in the vicinity of the mass and oscillation shells. In this approximation	 the evolution equation for the 
two-point correlator can be reduced to a closed-form evolution equation for the Wigner function. In its Liouville term the velocity
$\vec{v}$ is repla\-ced by the matrix of velocities, $\vec{v}_{ij}=2\vec{p}/(\omega_i+\omega_j)$. In the physically relevant 
relativistic limit  $\vec{v}_{ij}$ is well approximated by the average velocity  $\vec{\bar{v}}_{ij} = (\vec{v}_i+\vec{v}_j)/2$. In 
this approximation the form of the evolution equation for the Wigner function is identical to the form of the (collisionless 
limit of the) kinetic equation for the matrix of densities. The latter is derived using methods of perturbative quantum field theory.
Solutions of the evolution equation for the Wigner function, derived from the Schr\"odinger equation, can be 
obtained from known solutions of the latter. The resulting Wigner function is localized neither in the momentum, nor in the 
coordinate spaces. The stronger it peaks around a characteristic momentum  $\vec{p}_w$, the weaker is its dependence on 
$\vec{\bar{v}}t-\vec{x}$, and vice versa. In other words, the evolution equation for the Wigner 
function admits solutions consistent with the uncertainty principle. Let us emphasize that unlike the Schr\"odinger equation,
it does not enforce the uncertainty principle and admits also classical solutions describing a particle with 
definite coordinate and momentum. On the other hand, it consistently evolves solutions satisfying the uncertainty principle if
supplemented by respective initial conditions.

For a neutrino propagating in vacuum or collisionless matter each individual momentum mode of the Wigner function 
does not experience any suppression in the course of the neutrino propagation. On the other hand, their growing dephasing 
results in kinematical decoherence in the density matrix at late times.

One can expect that the three approaches to neutrino oscillations produce identical results for neutrino propagation in vacuum 
or an external potential (up to small differences related to different approximation schemes). Results of the present work indicate that  
this is indeed the case. Because the evolution equations for the two-point correlator and  Wigner function, derived from the 
Schr\"odinger equation, match the quantum kinetic and kinetic equation respectively, the two-point correlator and  Wigner function,
derived from the solution of the Schr\"odinger equation, solve the quantum kinetic and kinetic equation respectively. This 
observation hints, that the effect of wave packet separation can be accounted for also in the (quantum) kinetic approach to 
neutrino oscillations by considering initial conditions consistent with the uncertainty principle. Given that the (quantum) kinetic 
approach naturally incorporates scattering processes, this finding speaks in favor of using the (quantum)  kinetic description for 
the analysis of neutrino propagation in exploding supernovae where neutrino oscillations and collisions, as well as the effect 
of wave packet separation, might be equally important.

\acknowledgments

We acknowledge support by the Russian Science Foundation under the Grant No.18-72-10070.  The author would like to thank
Alexandra Dobrynina and Peter Millington for valuable comments and careful proof-reading of the manuscript.

\appendix

\section{\label{sec:covliouville}Covariant form of the Liouville equation}
Rewritten in terms of the four-coordinate $x^\mu=(t,\vec{x})$ and on-shell four momentum $p^\mu=(\omega(\vec{p}),\vec{p})$
the solution of the Schr\"odinger equation \eqref{eq:wavepacket} takes the form 
\begin{align}
\label{eq:wavepacketcov}
\psi_i(x^\mu)=\int\frac{d^3\vec{p}}{(2\pi)^3} \psi_i(0,\vec{p})e^{-ip_\mu x^\mu}\,.
\end{align}
Substituting equation \eqref{eq:wavepacketcov} into the definition of the two-point correlator, 
\begin{align}
\varrho_{ij}(x,p)=\int d^4y\, e^{ip_\mu y^\mu}\psi_i(x^\mu+y^\mu\!/2)\psi^*_j(x^\mu-y^\mu\!/2)\,,
\end{align}
we obtain 
\begin{align}
\varrho_{ij}(x,p)=\int\frac{d^3\vec{k}}{(2\pi)^3}\frac{d^3\vec{q}}{(2\pi)^3}\psi_i(0,\vec{k})\psi^*_j(0,\vec{q})
(2\pi)^4\delta\left(p-\frac{k+q}{2}\right) e^{-i(k_\nu-q_\nu)x^\nu}.
\end{align}
Here $k$ and $q$ are the on-shell four momenta of the respective mass eigenstates, $k^2=m^2_i$ and 
$q^2=m^2_j$. Action of the Liouville operator yields
\begin{align}
p^\mu\partial_\mu\varrho_{ij}(x,p)  =& -i\int\frac{d^3k}{(2\pi)^3}\frac{d^3q}{(2\pi)^3}\psi_i(0,\vec{k})\psi^*_j(0,\vec{q})
(2\pi)^4\delta\left(p-\frac{k+q}{2}\right) e^{-i(k_\nu-q_\nu)x^\nu}\nonumber\\
& \times p^\mu(k_\mu-q_\mu)\,.
\end{align} 
The delta-function further implies $p^\mu=(k^\mu+q^\mu)/2$. Thus 
$p^\mu(k_\mu-q_\mu)=(k^\mu+q^\mu)(k_\mu-q_\mu)/2=(k^2-q^2)/2=(m^2_i-m^2_j)/2$ and we recover 
\begin{align}
p^\mu\partial_\mu\varrho_{ij}(x,p) = -{\textstyle\frac{i}{2}}(m^2_i-m^2_j)\varrho_{ij}(x,p)\,.
\end{align}

\section{\label{sec:lorentzian}Lorentzian initial conditions}
To demonstrate that the Wigner function is consistent with the uncertainty principle also for Lorentzian initial conditions,
\begin{align}
\label{eq:lorenzianwavepacket}
\psi_i(0,{\rm p})\propto \frac{1}{1+({\rm p}-{\rm p}_w)^2/\sigma^2_p}\,,
\end{align}
it is sufficient to consider neutrino propagation in 1+1 dimensions. Substituting equation \eqref{eq:lorenzianwavepacket} 
into (the 1+1 dimensional version of) equation \eqref{eq:shapewigner} and integrating over $\Delta$ we arrive at 
\begin{align}
\label{eq:shapefactorlorentzian}
g_{ij}({\rm v}_{ij}&({\rm p})t-{\rm x},{\rm p}) \propto \frac{1}{1+({\rm p}-{\rm p}_w)^2/\sigma^2_p}
\exp\left(-\frac{|{\rm v}_{ij}({\rm p})t-{\rm x}|}{\sigma_x}\right)\nonumber\\
&\times \left[\frac{\sigma_p}{{\rm p}-{\rm p}_w} \sin\left(2({\rm p}-{\rm p}_w)|{\rm v}_{ij}({\rm p})t-{\rm x}|\right)
+\cos\left(2({\rm p}-{\rm p}_w)|{\rm v}_{ij}({\rm p})t-{\rm x}|\right)\right]\,.
\end{align}
As can be inferred from equation \eqref{eq:shapefactorlorentzian}, localization in the momentum space is inversely 
proportional to localization in the coordinate space, in agreement with the uncertainty principle.

\section{\label{sec:constantpot}Propagation in a constant potential}
In the momentum representation equation \eqref{eq:correlator} can be recast in the form 
\begin{align}
\label{eq:twopointcorrmomentumspace}
\varrho_{ij}(t,\vec{x},\epsilon,\vec{p}) =\int d\tau e^{i\epsilon\tau}\int \frac{d^3\vec{\Delta}}{(2\pi)^3}e^{i\vec{\Delta}\vec{x}}
\psi_i(t+\tau/2,\vec{p}+\vec{\Delta}/2)\psi^{*}_j(t-\tau/2,\vec{p}-\vec{\Delta}/2)\,.
\end{align}
Fourier-transforming equation \eqref{eq:schroedinger} we obtain the momentum-representation Schr\"odinger equation
\begin{align}
\label{eq:Schrodingermomrepr}
i\partial_t \psi_i(t,\vec{p}) = \int\frac{d^3\vec{k}}{(2\pi)^3} {\sf H}_{ij}(t,\vec{p},\vec{k}) \psi_j(t,\vec{k})\,.
\end{align}
The momentum-representation Hamiltonian is related to its coordinate-representation counterpart by
\begin{align}
\label{eq:Hamiltonianconstpot}
{\sf H}_{ij}(t,\vec{p},\vec{k}) = \int d^3 \vec{x}\, e^{-i\vec{p}\vec{x}}{\sf H}_{ij}(t,\vec{x})e^{i\vec{k}\vec{x}}\,.
\end{align}
Substituting  
${\sf H}_{ij}(\vec{x})=\delta_{ij}(-\partial^2_{\vec{x}}+m^2_i)^{\sfrac{1}{\,2}}+{\sf V}_{ij}$ 
we obtain 
${\sf H}_{ij}(t,\vec{p},\vec{k})=(2\pi)^3\delta(\vec{p}-\vec{k}){\sf H}_{ij}(\vec{p})$ with 
${\sf H}_{ij}(\vec{p})=\delta_{ij}\,\omega_i(\vec{p})+{\sf V}_{ij}$.
Equation \eqref{eq:Schrodingermomrepr} further implies 
\begin{align}
\label{eq:schrodeqconstpot} 
i\partial_t\psi_i(t,\vec{p})={\sf H}_{ij}(\vec{p}) \psi_j(t,\vec{p})\,.
\end{align}
The Hamiltonian is diagonalized by a unitary transformation,  $U^\dagger_{in}(\vec{p}){\sf H}_{nm}(\vec{p}) U_{mj}(\vec{p})=
\delta_{ij}\Omega_i(\vec{p})$. The wave function in the propagation basis, \smash{$\Psi_i(t,\vec{p})=U^\dagger_{ij}(t,\vec{p}) \psi_j(t,\vec{p})$}, satisfies the Schr\"odinger equation $i\partial_t\Psi_i(t,\vec{p}) = \Omega_i(\vec{p}) \Psi_i(t,\vec{p})$. Its solution reads 
$\Psi_i(t,\vec{p})=e^{-i\Omega_i(\vec{p})t}\Psi_i(0,\vec{p})$. Transforming the solution back to the neutrino mass basis we 
obtain 
\begin{align}
\label{eq:solShroedingereq}
\psi_i(t,\vec{p}) = U_{in}(\vec{p}) e^{-i\Omega_n(\vec{p})t} \Psi_n(0,\vec{p})\,.
\end{align}
Substituting equation \eqref{eq:solShroedingereq} into equation \eqref{eq:twopointcorrmomentumspace} and integrating 
over $\tau$ we find for the two-point correlator  
\begin{align}
\label{eq:twopointcorrconstpot}
\varrho_{ij}(t,\vec{x},\epsilon,\vec{p}) & = \int \frac{d^3\vec{\Delta}}{(2\pi)^3}e^{i\vec{\Delta}\vec{x}} 
\,2\pi\delta\left(\epsilon-\frac{\Omega_n(\vec{p}+\vec{\Delta}/2)+\Omega_m(\vec{p}-\vec{\Delta}/2)}{2}\right)\nonumber\\
&\times e^{-i(\Omega_n(\vec{p}+\vec{\Delta}/2)-\Omega_m(\vec{p}-\vec{\Delta}/2))t}\,
\nonumber\\
& \times  U_{in}(\vec{p}+\vec{\Delta}/2)\Psi_n(0,\vec{p}+\vec{\Delta}/2)\Psi^*_m(0,\vec{p}-\vec{\Delta}/2)
U^\dagger_{mj}(\vec{p}-\vec{\Delta}/2)\,.
\end{align}
As expected, the spectrum of the two-point correlator, $2\epsilon=\Omega_n(\vec{p}+\vec{\Delta}/2)+
\Omega_m(\vec{p}-\vec{\Delta}/2)$, is determined by eigenvalues of the Hamiltonian.  Proceeding as in section 
\ref{sec:liouvilleterm}  we represent the phase 
difference as 
\begin{align}
\Omega_n(\vec{p}+\vec{\Delta}/2)-\Omega_m(\vec{p}-\vec{\Delta}/2) = 
\frac{\Omega^2_n(\vec{p}+\vec{\Delta}/2)-\Omega^2_m(\vec{p}-\vec{\Delta}/2)}{2\epsilon}\,.
\end{align}
The resulting time derivative of the two-point correlator is given by
\begin{align}
\partial_t & \varrho_{ij}(t,\vec{x},\epsilon,\vec{p})  = -\frac{i}{2\epsilon}\int \frac{d^3\vec{\Delta}}{(2\pi)^3}e^{i\vec{\Delta}\vec{x}} 
\,2\pi\delta\left(\epsilon-\frac{\Omega_n(\vec{p}+\vec{\Delta}/2)+\Omega_m(\vec{p}-\vec{\Delta}/2)}{2}\right)\nonumber\\
&\times e^{-i(\Omega^2_n(\vec{p}+\vec{\Delta}/2)-\Omega^2_m(\vec{p}-\vec{\Delta}/2))t/2\epsilon}\,
\nonumber\\
& \times [U_{in}(\vec{p}+\vec{\Delta}/2)\Omega^2_n(\vec{p}+\vec{\Delta}/2)
\Psi_n(0,\vec{p}+\vec{\Delta}/2)\Psi^*_m(0,\vec{p}-\vec{\Delta}/2)U^\dagger_{mj}(\vec{p}-\vec{\Delta}/2)\nonumber\\
&- U_{in}(\vec{p}+\vec{\Delta}/2)\Psi_n(0,\vec{p}+\vec{\Delta}/2)\Psi^*_m(0,\vec{p}-\vec{\Delta}/2)
\Omega^2_m(\vec{p}-\vec{\Delta}/2)U^\dagger_{mj}(\vec{p}-\vec{\Delta}/2)]\,.
\end{align}
Inserting $\delta_{ln} = \sum_kU^\dagger_{lk}(\vec{p}+\vec{\Delta}/2)U_{kn}(\vec{p}+\vec{\Delta}/2)$ and 
$\delta_{ml} = \sum_kU^\dagger_{mk}(\vec{p}-\vec{\Delta}/2)U_{kl}(\vec{p}-\vec{\Delta}/2)$ we can recast it 
in the form 
\begin{align}
\partial_t & \varrho_{ij}(t,\vec{x},\epsilon,\vec{p})  = -\frac{i}{2\epsilon}\int \frac{d^3\vec{\Delta}}{(2\pi)^3}e^{i\vec{\Delta}\vec{x}} 
\,2\pi\delta\left(\epsilon-\frac{\Omega_n(\vec{p}+\vec{\Delta}/2)+\Omega_m(\vec{p}-\vec{\Delta}/2)}{2}\right)\nonumber\\
&\times e^{-i(\Omega^2_n(\vec{p}+\vec{\Delta}/2)-\Omega^2_m(\vec{p}-\vec{\Delta}/2))t/2\epsilon}\,
\nonumber\\
& \times [{\sf H}^2_{ik}(\vec{p}+\vec{\Delta}/2)U_{kn}(\vec{p}+\vec{\Delta}/2)
\Psi_n(0,\vec{p}+\vec{\Delta}/2)\Psi^*_m(0,\vec{p}-\vec{\Delta}/2)U^\dagger_{mj}(\vec{p}-\vec{\Delta}/2)\nonumber\\
&- U_{in}(\vec{p}+\vec{\Delta}/2)\Psi_n(0,\vec{p}+\vec{\Delta}/2)\Psi^*_m(0,\vec{p}-\vec{\Delta}/2)
U^\dagger_{mk}(\vec{p}-\vec{\Delta}/2) {\sf H}^2_{kj}(\vec{p}-\vec{\Delta}/2)]\,.
\end{align}
To the first order in the gradients ${\sf H}^2_{ij}(\vec{p}\pm\vec{\Delta}/2)\approx {\sf H}^2_{ij}(\vec{p})\pm 
\textstyle{\frac12} \partial_{\vec{p}}{\sf H}^2_{ij}(\vec{p})	\vec{\Delta}$.  Employing  $\vec{\Delta}e^{i\vec{\Delta}\vec{x}}=
-i\partial_{\vec{x}}e^{i\vec{\Delta}\vec{x}}$ and multiplying the left- and right-hand side by $\epsilon$ we finally recover equation \eqref{eq:liouvilleeqconstpot},
\begin{align}
\label{eq:liouvilleeqconstpotapp}
\epsilon\,\partial_t\varrho(t,\vec{x},\epsilon,\vec{p}) 
+ {\textstyle\frac14}\bigl\{\partial_{\vec{p}}{\sf H}^2(\vec{p}),\partial_{\vec{x}}\varrho(t,\vec{x},\epsilon,\vec{p})\bigr\}
= -{\textstyle\frac{i}{2}}\bigl[{\sf H}^2(\vec{p}),\varrho(t,\vec{x},\epsilon,\vec{p})\bigr]\,.
\end{align}
As can be read off from equation \eqref{eq:twopointcorrconstpot}, the spectrum of the two-point correlator is determined by 
the eigenvalues of the Hamiltonian and is shifted with respect to the vacuum one by terms proportional to the potential. 
Therefore, evolution equation \eqref{eq:liouvilleeqconstpotapp} must be supplemented by initial conditions that 
consistently take into account modifications of the neutrino spectrum in an external potential.  
 
\section{\label{sec:timedeppot}Propagation in a time-dependent potential}
The structure of the left-hand side of equation \eqref{eq:qktimedeppot} is not specific to mixing neutrinos. In order to better 
understand the origin of individual terms in equation \eqref{eq:qktimedeppot} we rederive it for a single neutrino generation
here.
 
Equations \eqref{eq:Schrodingermomrepr} and \eqref{eq:Hamiltonianconstpot} imply that for a homogeneous time-dependent 
potential the Schr\"odin\-ger equation in the momentum representation reads 
\begin{align}
\label{eq:schrodeqtimedeppot} 
i\partial_t\psi(t,\vec{p})={\sf H}(t,\vec{p}) \psi(t,\vec{p})\,,
\end{align}
with ${\sf H}(t,\vec{p})=\omega(\vec{p})+{\sf V}(t)$. Its solution is given by \cite{Hansen:2016klk} 
\begin{align}
\label{eq:schrodeqsol}
\psi(t,\vec{p}) = e^{-i\int^t_0 dt' {\sf H}(t'\!,\,\vec{p})}\,\psi(0,\vec{p})\,.
\end{align}
Substituting equation \eqref{eq:schrodeqsol} into equation \eqref{eq:twopointcorrmomentumspace} we obtain 
\begin{align}
\label{eq:timedeppotstep0}
\varrho(t,\vec{x},\epsilon,\vec{p})  & = \int d\tau e^{i\epsilon\tau}\int\frac{d^3\vec{\Delta}}{(2\pi)^3}	e^{i\vec{\Delta}\vec{x}}\,
\psi(0,\vec{p}+\vec{\Delta}/2)\psi^*(0,\vec{p}-\vec{\Delta}/2)\nonumber\\
&\times e^{-i(\omega(\vec{p}+\vec{\Delta}/2)-\omega(\vec{p}-\vec{\Delta}/2))t}
e^{-i(\omega(\vec{p}+\vec{\Delta}/2)+\omega(\vec{p}-\vec{\Delta}/2))\tau/2} e^{-i\int^{t+\tau/2}_{t-\tau/2}dt' {\sf V}(t')}\,.
\end{align}
Its time derivative is given by
\begin{align}
\label{eq:timedeppotstep1}
\partial_t \varrho(t,\vec{x},\epsilon,\vec{p})  & = -i\int d\tau e^{i\epsilon\tau}\int\frac{d^3\vec{\Delta}}{(2\pi)^3}	e^{i\vec{\Delta}\vec{x}}\,
\psi(0,\vec{p}+\vec{\Delta}/2)\psi^*(0,\vec{p}-\vec{\Delta}/2)\nonumber\\
&\times e^{-i(\omega(\vec{p}+\vec{\Delta}/2)-\omega(\vec{p}-\vec{\Delta}/2))t}
e^{-i(\omega(\vec{p}+\vec{\Delta}/2)+\omega(\vec{p}-\vec{\Delta}/2))\tau/2}e^{-i\int^{t+\tau/2}_{t-\tau/2}dt' {\sf V}(t')}\nonumber\\
&\times (\omega(\vec{p}+\vec{\Delta}/2)-\omega(\vec{p}-\vec{\Delta}/2)+\partial_t {\textstyle \int^{t+\tau/2}_{t-\tau/2}}dt' {\sf V}(t'))\,.
\end{align}
Multiplying equation \eqref{eq:timedeppotstep1} by $\epsilon$, using $\epsilon e^{i\epsilon \tau}=-i\partial_{\tau} e^{i\epsilon \tau}$, 
and further replacing $-i\partial_{\tau} e^{i\epsilon \tau}$ by $i e^{i\epsilon \tau} \partial_{\tau}$ we obtain upon integration by parts 
\begin{align}
\label{eq:timedeppotstep2}
\epsilon\partial_t \varrho(t,\vec{x},\epsilon,\vec{p})  & = \int d\tau e^{i\epsilon\tau}\int\frac{d^3\vec{\Delta}}{(2\pi)^3}	
e^{i\vec{\Delta}\vec{x}}\,
\psi(0,\vec{p}+\vec{\Delta}/2)\psi^*(0,\vec{p}-\vec{\Delta}/2)\nonumber\\
&\times e^{-i(\omega(\vec{p}+\vec{\Delta}/2)-\omega(\vec{p}-\vec{\Delta}/2))t}
e^{-i(\omega(\vec{p}+\vec{\Delta}/2)+\omega(\vec{p}-\vec{\Delta}/2))\tau/2}e^{-i\int^{t+\tau/2}_{t-\tau/2}dt' {\sf V}(t')}\nonumber\\
&\times \bigl[-{\textstyle\frac{i}{2}}(\omega(\vec{p}+\vec{\Delta}/2)-\omega(\vec{p}-\vec{\Delta}/2)+
\partial_t {\textstyle \int^{t+\tau/2}_{t-\tau/2}}dt' {\sf V}(t'))\nonumber\\
&\times (\omega(\vec{p}+\vec{\Delta}/2)+\omega(\vec{p}-\vec{\Delta}/2)+2\partial_\tau{\textstyle\int^{t+\tau/2}_{t-\tau/2}} dt' {\sf V}(t'))
\nonumber\\
&+ \partial_t\partial_\tau {\textstyle \int^{t+\tau/2}_{t-\tau/2}}dt' {\sf V}(t')\bigr]\,.
\end{align}
Using $\partial_t\! \int^{t+\tau/2}_{t-\tau/2} dt' {\sf V}(t')={\sf V}(t+\tau/2)-{\sf V}(t-\tau/2)$ and  
$2\partial_\tau\!	 \int^{t+\tau/2}_{t-\tau/2} dt' {\sf V}(t')={\sf V}(t+\tau/2)+{\sf V}(t-\tau/2)$ we can rewrite equation 
\eqref{eq:timedeppotstep2} in the form 
\begin{align}
\label{eq:timedeppotstep3}
\epsilon\partial_t \varrho(t,\vec{x},\epsilon,\vec{p})  & =
\frac12\int d\tau e^{i\epsilon\tau}\int\frac{d^3\vec{\Delta}}{(2\pi)^3}	e^{i\vec{\Delta}\vec{x}}\,
\psi(0,\vec{p}+\vec{\Delta}/2)\psi^*(0,\vec{p}-\vec{\Delta}/2)\nonumber\\
&\times e^{-i(\omega(\vec{p}+\vec{\Delta}/2)-\omega(\vec{p}-\vec{\Delta}/2))t}
e^{-i(\omega(\vec{p}+\vec{\Delta}/2)+\omega(\vec{p}-\vec{\Delta}/2))\tau/2}e^{-i\int^{t+\tau/2}_{t-\tau/2}dt' {\sf V}(t')}\nonumber\\
&\times \bigl[-i({\sf H}^2(t+\tau/2,\vec{p}+\vec{\Delta}/2)-{\sf H}^2(t-\tau/2,\vec{p}-\vec{\Delta}/2))
\nonumber\\
&+(\partial_t{\sf V}(t+\tau/2)+\partial_t {\sf V}(t-\tau/2))\bigr]\,,
\end{align}
thereby essentially recovering the structure of equations \eqref{eq:qktimedeppotstep1}  and \eqref{eq:qktimedeppotstep2}.
To the first order in the gradient expansion ${\sf H}^2(t\pm\tau/2,\vec{p}\pm\vec{\Delta}/2)\approx {\sf H}^2(t,\vec{p})\pm
\partial_\vec{p}{\sf H}^2(t,\vec{p})\vec{\Delta}/2\pm\partial_t{\sf H}^2(t,\vec{p})\tau/2$ and similarly
$\partial_t{\sf V}(t+\tau/2)+\partial_t {\sf V}(t-\tau/2)\approx 2\partial_t{\sf V}(t)$. Using $e^{i\vec{\Delta}\vec{x}}\vec{\Delta}
=-i\partial_\vec{x}e^{i\vec{\Delta}\vec{x}}$ and $e^{i\epsilon\tau}\tau=-i\partial_\epsilon e^{i\epsilon\tau}$ we obtain 
from equation \eqref{eq:timedeppotstep3}
\begin{align}
\label{eq:qktimedeppotapp}
\epsilon\partial_t\varrho(t,\vec{x},\epsilon,\vec{p})
&+{\textstyle\frac12}\partial_\vec{p}{\sf H}^2(t,\vec{p})\partial_\vec{x}\varrho(t,\vec{x},\epsilon,\vec{p})
+{\textstyle\frac12}\partial_t{\sf H}^2(t,\vec{p})\partial_\epsilon\varrho(t,\vec{x},\epsilon,\vec{p})\nonumber\\
&- \partial_t{\sf V}(t)\varrho(t,\vec{x},\epsilon,\vec{p}) \approx 0\,,
\end{align}
which (for a single neutrino generation) reproduces equation \eqref{eq:qktimedeppot}.

In fact, a partial gradient expansion can be performed directly in equation \eqref{eq:timedeppotstep0}. Approximating 
$\int^{t+\tau/2}_{t-\tau/2}dt' {\sf V}(t')$ by ${\sf V}(t)\tau$ we obtain
\begin{align}
\label{eq:timedeppotstep4}
\varrho(t,\vec{x},\epsilon,\vec{p})  & = \int d\tau e^{i\epsilon\tau}\int\frac{d^3\vec{\Delta}}{(2\pi)^3}	e^{i\vec{\Delta}\vec{x}}\,
\psi(0,\vec{p}+\vec{\Delta}/2)\psi^*(0,\vec{p}-\vec{\Delta}/2)\nonumber\\
&\times e^{-i(\omega(\vec{p}+\vec{\Delta}/2)-\omega(\vec{p}-\vec{\Delta}/2))t}
e^{-i(\omega(\vec{p}+\vec{\Delta}/2)+\omega(\vec{p}-\vec{\Delta}/2))\tau/2} e^{-i{\sf V}(t)\tau}\,.
\end{align}
Performing steps \eqref{eq:timedeppotstep1} to \eqref{eq:timedeppotstep3} we would again arrive into equation 
\eqref{eq:qktimedeppotapp}. That is, the approximate expression equation \eqref{eq:timedeppotstep4}
also satisfies the same evolution equation. Integrating over $\tau$ in equation \eqref{eq:timedeppotstep4} 
we arrive at 
\begin{align}
\label{eq:timedeppotstep5}
\varrho(t,\vec{x},\epsilon,\vec{p})  & = \int\frac{d^3\vec{\Delta}}{(2\pi)^3}e^{i\vec{\Delta}\vec{x}}\,
\psi(0,\vec{p}+\vec{\Delta}/2)\psi^*(0,\vec{p}-\vec{\Delta}/2)\, 
e^{-i(\omega(\vec{p}+\vec{\Delta}/2)-\omega(\vec{p}-\vec{\Delta}/2))t}\nonumber\\
&\times 2\pi\,\delta\left(\epsilon-\frac{\omega(\vec{p}+\vec{\Delta}/2)+\omega(\vec{p}-\vec{\Delta}/2)+2{\sf V}(t)}{2}\right)\,.
\end{align}
As can be read off from equation \eqref{eq:timedeppotstep5}, a homogeneous time-dependent potential affects primarily the spectrum 
of the two-point correlator. Its time derivative is a sum of two contributions: the first contribution is proportional to the time derivative 
of the phase exponent while the second contribution is proportional to the time derivative of the energy-conserving delta-function. 

The first contribution reads
\begin{align}
\label{eq:timedeppotstep6}
\epsilon\partial_t\varrho(t,\vec{x},\epsilon,\vec{p})  & \in -i\int\frac{d^3\vec{\Delta}}{(2\pi)^3}e^{i\vec{\Delta}\vec{x}}\,
\psi(0,\vec{p}+\vec{\Delta}/2)\psi^*(0,\vec{p}-\vec{\Delta}/2)\, 
e^{-i(\omega(\vec{p}+\vec{\Delta}/2)-\omega(\vec{p}-\vec{\Delta}/2))t}\nonumber\\
&\times\epsilon\cdot (\omega(\vec{p}+\vec{\Delta}/2)-\omega(\vec{p}-\vec{\Delta}/2))\nonumber\\
&\times 2\pi\,\delta\left(\epsilon-\frac{\omega(\vec{p}+\vec{\Delta}/2)+\omega(\vec{p}-\vec{\Delta}/2)+2{\sf V}(t)}{2}\right)\,.
\end{align}
On the right-hand side of equation \eqref{eq:timedeppotstep6} $\epsilon$ can be replaced by the value implied by the 
energy-con\-serving delta-function. To the first order in the gradient expansion this results in
$\epsilon\cdot (\omega(\vec{p}+\vec{\Delta}/2)-\omega(\vec{p}-\vec{\Delta}/2))\approx \frac12\partial_\vec{p}{\sf H}^2(t,\vec{p})\vec{\Delta}$. Further using $e^{i\vec{\Delta}\vec{x}}\vec{\Delta}=-i\partial_\vec{x}e^{i\vec{\Delta}\vec{x}}$ we recover the second 
term on the left-hand side of equation \eqref{eq:qktimedeppotapp}. 

Using the chain rule we can trade the time-derivative for the derivative with respect to $\epsilon$ in the second contribution. 
This yields 
\begin{align}
\label{eq:timedeppotstep7}
\epsilon\partial_t\varrho(t,\vec{x},\epsilon,\vec{p})  & \in \int\frac{d^3\vec{\Delta}}{(2\pi)^3}e^{i\vec{\Delta}\vec{x}}\,
\psi(0,\vec{p}+\vec{\Delta}/2)\psi^*(0,\vec{p}-\vec{\Delta}/2)\, 
e^{-i(\omega(\vec{p}+\vec{\Delta}/2)-\omega(\vec{p}-\vec{\Delta}/2))t}\nonumber\\
&\times 2\pi\,\epsilon\cdot \partial_\epsilon 	\delta\left(\epsilon-\frac{\omega(\vec{p}+\vec{\Delta}/2)+
\omega(\vec{p}-\vec{\Delta}/2)+2{\sf V}(t)}{2}\right)\cdot\partial_t{\sf V}(t)\,.
\end{align}
Using $\epsilon\partial_\epsilon f(\epsilon) = \partial_\epsilon(\epsilon f(\epsilon))-f(\epsilon)$  we can in turn recast equation
\eqref{eq:timedeppotstep7} as a sum of two terms. The first term that stems from equation \eqref{eq:timedeppotstep7} reads
\begin{align}
\label{eq:timedeppotstep8}
\epsilon\partial_t\varrho(t,\vec{x},\epsilon,\vec{p})  & \in \partial_\epsilon \int\frac{d^3\vec{\Delta}}{(2\pi)^3}e^{i\vec{\Delta}\vec{x}}\,
\psi(0,\vec{p}+\vec{\Delta}/2)\psi^*(0,\vec{p}-\vec{\Delta}/2)\, 
e^{-i(\omega(\vec{p}+\vec{\Delta}/2)-\omega(\vec{p}-\vec{\Delta}/2))t}\nonumber\\
&\times 2\pi\,\delta\left(\epsilon-\frac{\omega(\vec{p}+\vec{\Delta}/2)+
\omega(\vec{p}-\vec{\Delta}/2)+2{\sf V}(t)}{2}\right)\cdot\epsilon\partial_t{\sf V}(t)\,.
\end{align}
Replacing $\epsilon$ by the value implied by the energy-conserving delta-function on the right-hand side of equation 
\eqref{eq:timedeppotstep8} we obtain, to the first order in the gradient expansion, $\epsilon\partial_t{\sf V}(t) \approx
\frac12\partial_t{\sf H}^2(t,\vec{p})$. Thus, we recover the second term on the left-hand side of equation 
\eqref{eq:qktimedeppotapp}. The second term originating from equation \eqref{eq:timedeppotstep7} reads
\begin{align}
\label{eq:timedeppotstep9}
\epsilon\partial_t\varrho(t,\vec{x},\epsilon,\vec{p})  & \in  \int\frac{d^3\vec{\Delta}}{(2\pi)^3}e^{i\vec{\Delta}\vec{x}}\,
\psi(0,\vec{p}+\vec{\Delta}/2)\psi^*(0,\vec{p}-\vec{\Delta}/2)\, 
e^{-i(\omega(\vec{p}+\vec{\Delta}/2)-\omega(\vec{p}-\vec{\Delta}/2))t}\nonumber\\
&\times 2\pi\,\delta\left(\epsilon-\frac{\omega(\vec{p}+\vec{\Delta}/2)+
\omega(\vec{p}-\vec{\Delta}/2)+2{\sf V}(t)}{2}\right)\cdot \partial_t{\sf V}(t)\,.
\end{align}
It reproduces the last term  on the left-hand side of equation \eqref{eq:qktimedeppotapp}. Let us emphasize here that 
the two terms, $\partial_t{\sf H}^2(t,\vec{p})\partial_\epsilon\varrho(t,\vec{x},\epsilon,\vec{p})$ and 
$\partial_t{\sf V}(t)\varrho(t,\vec{x},\epsilon,\vec{p})$, have a common origin and stem from the time derivative of the 
energy-conserving delta-function. Hence, they both describe shifts of the spectrum induced by the time-dependent potential.

\section{\label{sec:quantumkineticeq}Quantum kinetic equation}
In the collisionless approximation (i.e. including only forward scattering) the quantum kinetic equation for the Wightman 
propagators, $\varrho_\lessgtr(u,v)$, of a system of mixing scalar fields reads \citep{Hohenegger:2008zk,Garny:2009qn}:
\begin{align}
\label{eq:qkeqforwghtmn1}
[\square_u+m^2 + \Sigma(u)]\varrho_\lessgtr(u,v) = 0\,.
\end{align}
Using the property $\varrho_\lessgtr(u,v) = \varrho^\dagger_\lessgtr(v,u)$, hermitian-conjugating the resulting equation,
and taking into account hermiticity of $\Sigma$ we can rewrite equation \eqref{eq:qkeqforwghtmn1} in the form 
$\varrho_\lessgtr(v,u)[\square_u+m^2 + \Sigma(u)] = 0$. Renaming $u \leftrightarrow v$ we obtain 
\begin{align}
\label{eq:qkeqforwghtmn2}
\varrho_\lessgtr(u,v)[\square_v+m^2 + \Sigma(v)] = 0\,.
\end{align}

Next, we subtract equation \eqref{eq:qkeqforwghtmn2} from equation \eqref{eq:qkeqforwghtmn1} and 
express $u$ and $v$ in terms of the center and relative coordinates, $x=(u+v)/2$ and $s=u-v$, in the resulting equation. 
Further expanding the self-energy to the first-order in the gradients we obtain:
\begin{align}
\label{eq:qkeqcoordspace}
\partial_s\partial_x \varrho_<(x,s) + {\textstyle\frac14}\{s\,\partial_x \Sigma(x),\varrho_<(x,s)\}
\approx - {\textstyle\frac12}[m^2+\Sigma(x),\varrho_<(x,s)]\,.
\end{align}
Note that following the standard convention we use the same symbol for the Wightman propagator considered as a function 
of the center and relative coordinates. Expressing $\varrho_<(x,s)$ in terms of its Wigner-transform,
\begin{align}
\varrho_<(x,s) = \int\frac{d^4p}{(2\pi)^4}e^{-ips}\varrho_<(x,p)\,,
\end{align}
and using $se^{-ips}\rightarrow i\partial_p e^{-ips} \rightarrow -ie^{-ips}\partial_p$ we obtain from equation 
\eqref{eq:qkeqcoordspace}:
\begin{align}
\label{eq:qkeqwignspace}
p\partial_x \varrho_<(x,p) + {\textstyle\frac14} \{\partial_x \Sigma(x),\partial_p\varrho_<(x,p)\}
\approx - {\textstyle\frac{i}2}[m^2+\Sigma(x),\varrho_<(x,p)]\,,
\end{align}
where $x=(t,\vec{x})$ and $p=(\epsilon,\vec{p})$. The Wightman propagator $\varrho_<(t,\vec{x},\epsilon,\vec{p})$ 
is the quantum kinetic counterpart of the quantum-mechanical two-point correlator $\varrho(t,\vec{x},\epsilon,\vec{p})$.

To rewrite equation \eqref{eq:qkeqwignspace} in a form analogous to equation \eqref{eq:qkeqgenericpot} we separate 
the time and space components of $\partial_x$. Introducing an effective Hamiltonian,
\begin{align}
{\sf H}^2(t,\vec{x},\vec{p}) \equiv \vec{p}^2 + m^2 + \Sigma(t,\vec{x}) = \omega^2(\vec{p}) + \Sigma(t,\vec{x})\,,
\end{align}
and using the identity 
${\textstyle\frac14}\{\partial_\vec{p}{\sf H}^2(t,\vec{x},\vec{p}),\partial_\vec{x}\varrho_<(t,\vec{x},\epsilon,\vec{p})\}=
\vec{p}\partial_\vec{x}\varrho_<(t,\vec{x},\epsilon,\vec{p})$ we can finally recast equation \eqref{eq:qkeqwignspace}
in the form:
\begin{align}
\epsilon\partial_t \varrho_<(t,\vec{x},\epsilon,\vec{p}) & 	+ 
\textstyle{\frac14}\{\partial_\vec{p}{\sf H}^2(t,\vec{x},\vec{p}),\partial_\vec{x}\varrho_<(t,\vec{x},\epsilon,\vec{p})\} - 
\textstyle{\frac14}\{\partial_\vec{x}{\sf H}^2(t,\vec{x},\vec{p}),\partial_\vec{p}\varrho_<(t,\vec{x},\epsilon,\vec{p})\} \nonumber\\
& +\textstyle{\frac14}\{\partial_t{\sf H}^2(t,\vec{x},\vec{p}),\partial_\epsilon \varrho_<(t,\vec{x},\epsilon,\vec{p})\}
\approx - {\textstyle\frac{i}2}[{\sf H}^2(t,\vec{x},\vec{p}),\varrho_<(t,\vec{x},\epsilon,\vec{p})]\,.
\end{align}


\begin{thebibliography}{10}

\bibitem{Davis:1968cp}
R.~Davis, Jr., D.~S. Harmer, and K.~C. Hoffman, {\it {Search for neutrinos from
  the sun}},  {\em Phys. Rev. Lett.} {\bf 20} (1968) 1205--1209.

\bibitem{Gribov:1968kq}
V.~N. Gribov and B.~Pontecorvo, {\it {Neutrino astronomy and lepton charge}},
  {\em Phys. Lett.} {\bf 28B} (1969) 493.

\bibitem{Gavrin:2007wc}
V.~N. Gavrin and B.~T. Cleveland, {\it {Radiochemical solar neutrino
  experiments}},  {\em Nucl. Phys. Proc. Suppl.} {\bf 221} (2011) 90--97,
  [\href{http://arxiv.org/abs/nucl-ex/0703012}{{\tt nucl-ex/0703012}}].

\bibitem{Strumia:2006db}
A.~Strumia and F.~Vissani, {\it {Neutrino masses and mixings and...}},
  \href{http://arxiv.org/abs/hep-ph/0606054}{{\tt hep-ph/0606054}}.

\bibitem{Hosaka:2005um}
J.~Hosaka et~al., {\it {Solar neutrino measurements in super-Kamiokande-I}},
  {\em Phys. Rev.} {\bf D73} (2006) 112001,
  [\href{http://arxiv.org/abs/hep-ex/0508053}{{\tt hep-ex/0508053}}].

\bibitem{Akhmedov:2009rb}
E.~K. Akhmedov and A.~{\relax Yu}. Smirnov, {\it {Paradoxes of neutrino
  oscillations}},  {\em Phys. Atom. Nucl.} {\bf 72} (2009) 1363--1381,
  [\href{http://arxiv.org/abs/0905.1903}{{\tt arXiv:0905.1903}}].

\bibitem{Hansen:2016klk}
R.~S.~L. Hansen and A.~{\relax Yu}. Smirnov, {\it {The Liouville equation for
  flavour evolution of neutrinos and neutrino wave packets}},  {\em JCAP} {\bf
  1612} (2016), no.~12 019, [\href{http://arxiv.org/abs/1610.00910}{{\tt
  arXiv:1610.00910}}].

\bibitem{Akhmedov:2017mcc}
E.~Akhmedov, J.~Kopp, and M.~Lindner, {\it {Collective neutrino oscillations
  and neutrino wave packets}},  {\em JCAP} {\bf 1709} (2017), no.~09 017,
  [\href{http://arxiv.org/abs/1702.08338}{{\tt arXiv:1702.08338}}].

\bibitem{Sigl:1992fn}
G.~Sigl and G.~Raffelt, {\it {General kinetic description of relativistic mixed
  neutrinos}},  {\em Nucl. Phys.} {\bf B406} (1993) 423--451.

\bibitem{Dolgov:1980cq}
A.~D. Dolgov, {\it {Neutrinos in the Early Universe}},  {\em Sov. J. Nucl.
  Phys.} {\bf 33} (1981) 700--706. [Yad. Fiz.33,1309(1981)].

\bibitem{Yamada:2000za}
S.~Yamada, {\it {Boltzmann equations for neutrinos with flavor mixings}},  {\em
  Phys. Rev.} {\bf D62} (2000) 093026,
  [\href{http://arxiv.org/abs/astro-ph/0002502}{{\tt astro-ph/0002502}}].

\bibitem{Vlasenko:2013fja}
A.~Vlasenko, G.~M. Fuller, and V.~Cirigliano, {\it {Neutrino Quantum
  Kinetics}},  {\em Phys. Rev.} {\bf D89} (2014), no.~10 105004,
  [\href{http://arxiv.org/abs/1309.2628}{{\tt arXiv:1309.2628}}].

\bibitem{Cirigliano:2014aoa}
V.~Cirigliano, G.~M. Fuller, and A.~Vlasenko, {\it {A New Spin on Neutrino
  Quantum Kinetics}},  {\em Phys. Lett.} {\bf B747} (2015) 27--35,
  [\href{http://arxiv.org/abs/1406.5558}{{\tt arXiv:1406.5558}}].

\bibitem{Vlasenko:2014bva}
A.~Vlasenko, G.~M. Fuller, and V.~Cirigliano, {\it {Prospects for
  Neutrino-Antineutrino Transformation in Astrophysical Environments}},
  \href{http://arxiv.org/abs/1406.6724}{{\tt arXiv:1406.6724}}.

\bibitem{Blaschke:2016xxt}
D.~N. Blaschke and V.~Cirigliano, {\it {Neutrino Quantum Kinetic Equations: The
  Collision Term}},  {\em Phys. Rev.} {\bf D94} (2016), no.~3 033009,
  [\href{http://arxiv.org/abs/1605.09383}{{\tt arXiv:1605.09383}}].

\bibitem{Richers:2019grc}
S.~A. Richers, G.~C. McLaughlin, J.~P. Kneller, and A.~Vlasenko, {\it {Neutrino
  Quantum Kinetics in Compact Objects}},
  \href{http://arxiv.org/abs/1903.00022}{{\tt arXiv:1903.00022}}.

\bibitem{Akhmedov:2010ms}
E.~K. Akhmedov and J.~Kopp, {\it {Neutrino oscillations: Quantum mechanics vs.
  quantum field theory}},  {\em JHEP} {\bf 04} (2010) 008,
  [\href{http://arxiv.org/abs/1001.4815}{{\tt arXiv:1001.4815}}]. [Erratum:
  JHEP10,052(2013)].

\bibitem{Cozzella:2018zwm}
G.~Cozzella and C.~Giunti, {\it {Mixed states for mixing neutrinos}},
  \href{http://arxiv.org/abs/1804.00184}{{\tt arXiv:1804.00184}}.

\bibitem{Grimus:2019hlq}
W.~Grimus, {\it {Revisiting the quantum field theory of neutrino oscillations
  in vacuum}},  \href{http://arxiv.org/abs/1910.13776}{{\tt arXiv:1910.13776}}.

\bibitem{Grimus:2003es}
W.~Grimus, {\it {Neutrino physics - Theory}},  {\em Lect. Notes Phys.} {\bf
  629} (2004) 169--214, [\href{http://arxiv.org/abs/hep-ph/0307149}{{\tt
  hep-ph/0307149}}].

\bibitem{Grimus:1998uh}
W.~Grimus, P.~Stockinger, and S.~Mohanty, {\it {The Field theoretical approach
  to coherence in neutrino oscillations}},  {\em Phys. Rev.} {\bf D59} (1999)
  013011, [\href{http://arxiv.org/abs/hep-ph/9807442}{{\tt hep-ph/9807442}}].

\bibitem{Grimus:1996av}
W.~Grimus and P.~Stockinger, {\it {Real oscillations of virtual neutrinos}},
  {\em Phys. Rev.} {\bf D54} (1996) 3414--3419,
  [\href{http://arxiv.org/abs/hep-ph/9603430}{{\tt hep-ph/9603430}}].

\bibitem{Stirner:2018ojk}
T.~Stirner, G.~Sigl, and G.~Raffelt, {\it {Liouville term for neutrinos: Flavor
  structure and wave interpretation}},  {\em JCAP} {\bf 1805} (2018), no.~05
  016, [\href{http://arxiv.org/abs/1803.04693}{{\tt arXiv:1803.04693}}].

\bibitem{Sirera:1998ia}
M.~Sirera and A.~Perez, {\it {Relativistic Wigner function approach to neutrino
  propagation in matter}},  {\em Phys. Rev.} {\bf D59} (1999) 125011,
  [\href{http://arxiv.org/abs/hep-ph/9810347}{{\tt hep-ph/9810347}}].

\bibitem{Cardall:2007zw}
C.~Y. Cardall, {\it {Liouville equations for neutrino distribution matrices}},
  {\em Phys. Rev.} {\bf D78} (2008) 085017,
  [\href{http://arxiv.org/abs/0712.1188}{{\tt arXiv:0712.1188}}].

\bibitem{Garny:2011hg}
M.~Garny, A.~Kartavtsev, and A.~Hohenegger, {\it {Leptogenesis from first
  principles in the resonant regime}},  {\em Annals Phys.} {\bf 328} (2013)
  26--63, [\href{http://arxiv.org/abs/1112.6428}{{\tt arXiv:1112.6428}}].

\bibitem{Kartavtsev:2015vto}
A.~Kartavtsev, P.~Millington, and H.~Vogel, {\it {Lepton asymmetry from mixing
  and oscillations}},  {\em JHEP} {\bf 06} (2016) 066,
  [\href{http://arxiv.org/abs/1601.03086}{{\tt arXiv:1601.03086}}].

\bibitem{Herranen:2008di}
M.~Herranen, K.~Kainulainen, and P.~M. Rahkila, {\it {Kinetic theory for scalar
  fields with nonlocal quantum coherence}},  {\em JHEP} {\bf 0905} (2009) 119,
  [\href{http://arxiv.org/abs/0812.4029}{{\tt arXiv:0812.4029}}].

\bibitem{Herranen:2010mh}
M.~Herranen, K.~Kainulainen, and P.~M. Rahkila, {\it {Coherent quantum
  Boltzmann equations from cQPA}},  {\em JHEP} {\bf 12} (2010) 072,
  [\href{http://arxiv.org/abs/1006.1929}{{\tt arXiv:1006.1929}}].

\bibitem{Herranen:2011zg}
M.~Herranen, K.~Kainulainen, and P.~M. Rahkila, {\it {Flavour-coherent
  propagators and Feynman rules: Covariant cQPA formulation}},  {\em JHEP} {\bf
  02} (2012) 080, [\href{http://arxiv.org/abs/1108.2371}{{\tt
  arXiv:1108.2371}}].

\bibitem{Fidler:2011yq}
C.~Fidler, M.~Herranen, K.~Kainulainen, and P.~M. Rahkila, {\it {Flavoured
  quantum Boltzmann equations from cQPA}},  {\em JHEP} {\bf 1202} (2012) 065,
  [\href{http://arxiv.org/abs/1108.2309}{{\tt arXiv:1108.2309}}].

\bibitem{Blasone:1998hf}
M.~Blasone, P.~A. Henning, and G.~Vitiello, {\it {The Exact formula for
  neutrino oscillations}},  {\em Phys. Lett.} {\bf B451} (1999) 140--145,
  [\href{http://arxiv.org/abs/hep-th/9803157}{{\tt hep-th/9803157}}].

\bibitem{Bernardini:2004wr}
A.~E. Bernardini and S.~De~Leo, {\it {Dirac spinors and flavor oscillations}},
  {\em Eur. Phys. J.} {\bf C37} (2004) 471--480,
  [\href{http://arxiv.org/abs/hep-ph/0411153}{{\tt hep-ph/0411153}}].

\bibitem{Blasone:2018ktu}
M.~Blasone, P.~Jizba, and L.~Smaldone, {\it {Flavor Energy uncertainty
  relations for neutrino oscillations in quantum field theory}},  {\em Phys.
  Rev.} {\bf D99} (2019), no.~1 016014,
  [\href{http://arxiv.org/abs/1810.01648}{{\tt arXiv:1810.01648}}].

\bibitem{Blasone:2019rxl}
M.~Blasone, L.~Smaldone, and G.~Vitiello, {\it {Flavor neutrino states for
  pedestrians}},  \href{http://arxiv.org/abs/1903.01401}{{\tt
  arXiv:1903.01401}}.

\bibitem{Kobach:2017osm}
A.~Kobach, A.~V. Manohar, and J.~McGreevy, {\it {Neutrino Oscillation
  Measurements Computed in Quantum Field Theory}},  {\em Phys. Lett.} {\bf
  B783} (2018) 59--75, [\href{http://arxiv.org/abs/1711.07491}{{\tt
  arXiv:1711.07491}}].

\bibitem{Dickinson:2016oiy}
R.~Dickinson, J.~Forshaw, and P.~Millington, {\it {Probabilities and signalling
  in quantum field theory}},  {\em Phys. Rev.} {\bf D93} (2016), no.~6 065054,
  [\href{http://arxiv.org/abs/1601.07784}{{\tt arXiv:1601.07784}}].

\bibitem{Ferretti:1968}
B.~Ferretti, {\em Old and New Problems in Elementary Particles}, p.~108.
\newblock Academic Press.
\newblock New York, 1968.

\bibitem{PhysRevA.56.3395}
E.~A. Power and T.~Thirunamachandran, {\it Analysis of the causal behavior in
  energy transfer between atoms},  {\em Phys. Rev. A} {\bf 56} (Nov, 1997)
  3395--3408.

\bibitem{Raffelt:2010za}
G.~G. Raffelt and I.~Tamborra, {\it {Synchronization versus decoherence of
  neutrino oscillations at intermediate densities}},  {\em Phys. Rev.} {\bf
  D82} (2010) 125004, [\href{http://arxiv.org/abs/1006.0002}{{\tt
  arXiv:1006.0002}}].

\bibitem{Airen:2018nvp}
S.~Airen, F.~Capozzi, S.~Chakraborty, B.~Dasgupta, G.~Raffelt, and T.~Stirner,
  {\it {Normal-mode Analysis for Collective Neutrino Oscillations}},  {\em
  JCAP} {\bf 1812} (2018), no.~12 019,
  [\href{http://arxiv.org/abs/1809.09137}{{\tt arXiv:1809.09137}}].

\bibitem{Capozzi:2018clo}
F.~Capozzi, B.~Dasgupta, A.~Mirizzi, M.~Sen, and G.~Sigl, {\it {Collisional
  triggering of fast flavor conversions of supernova neutrinos}},  {\em Phys.
  Rev. Lett.} {\bf 122} (2019), no.~9 091101,
  [\href{http://arxiv.org/abs/1808.06618}{{\tt arXiv:1808.06618}}].

\bibitem{Capozzi:2019lso}
F.~Capozzi, G.~Raffelt, and T.~Stirner, {\it {Fast Neutrino Flavor Conversion:
  Collective Motion vs. Decoherence}},
  \href{http://arxiv.org/abs/1906.08794}{{\tt arXiv:1906.08794}}.

\bibitem{Abbar:2019zoq}
S.~Abbar, H.~Duan, K.~Sumiyoshi, T.~Takiwaki, and M.~C. Volpe, {\it {Fast
  Neutrino Flavor Conversion Modes in Multidimensional Core-collapse Supernova
  Models: the Role of the Asymmetric Neutrino Distributions}},
  \href{http://arxiv.org/abs/1911.01983}{{\tt arXiv:1911.01983}}.

\bibitem{Johns:2019izj}
L.~Johns, H.~Nagakura, G.~M. Fuller, and A.~Burrows, {\it {Neutrino
  oscillations in supernovae: angular moments and fast instabilities}},
  \href{http://arxiv.org/abs/1910.05682}{{\tt arXiv:1910.05682}}.

\bibitem{Chakraborty:2019wxe}
M.~Chakraborty and S.~Chakraborty, {\it {Three flavor neutrino conversions in
  supernovae: Slow $\&$ Fast instabilities}},
  \href{http://arxiv.org/abs/1909.10420}{{\tt arXiv:1909.10420}}.

\bibitem{Cervia:2019res}
M.~J. Cervia, A.~V. Patwardhan, A.~B. Balantekin, S.~N. Coppersmith, and C.~W.
  Johnson, {\it {Entanglement and collective flavor oscillations in a dense
  neutrino gas}},  {\em Phys. Rev.} {\bf D100} (2019), no.~8 083001,
  [\href{http://arxiv.org/abs/1908.03511}{{\tt arXiv:1908.03511}}].

\bibitem{Doring:2019axc}
C.~Döring, R.~S.~L. Hansen, and M.~Lindner, {\it {Stability of three neutrino
  flavor conversion in supernovae}},  {\em JCAP} {\bf 1908} (2019) 003,
  [\href{http://arxiv.org/abs/1905.03647}{{\tt arXiv:1905.03647}}].

\bibitem{Shalgar:2019qwg}
S.~Shalgar, I.~Padilla-Gay, and I.~Tamborra, {\it {Neutrino propagation hinders
  fast pairwise flavor conversions}},
  \href{http://arxiv.org/abs/1911.09110}{{\tt arXiv:1911.09110}}.

\bibitem{Kersten:2015kio}
J.~Kersten and A.~{\relax Yu}. Smirnov, {\it {Decoherence and oscillations of
  supernova neutrinos}},  {\em Eur. Phys. J.} {\bf C76} (2016), no.~6 339,
  [\href{http://arxiv.org/abs/1512.09068}{{\tt arXiv:1512.09068}}].

\bibitem{Lindblad:1975ef}
G.~Lindblad, {\it {On the Generators of Quantum Dynamical Semigroups}},  {\em
  Commun. Math. Phys.} {\bf 48} (1976) 119.

\bibitem{Benatti:2000ph}
F.~Benatti and R.~Floreanini, {\it {Open system approach to neutrino
  oscillations}},  {\em JHEP} {\bf 02} (2000) 032,
  [\href{http://arxiv.org/abs/hep-ph/0002221}{{\tt hep-ph/0002221}}].

\bibitem{PhysRevA.45.2243}
M.~J. Thomson, {\it Damping of quantum coherence by elastic and inelastic
  processes},  {\em Phys. Rev. A} {\bf 45} (Feb, 1992) 2243--2249.

\bibitem{Mikheyev:1989dy}
S.~P. Mikheyev and A.~{\relax Yu}. Smirnov, {\it {Resonant neutrino
  oscillations in matter}},  {\em Prog. Part. Nucl. Phys.} {\bf 23} (1989)
  41--136.

\bibitem{Maltoni:2015kca}
M.~Maltoni and A.~{\relax Yu}. Smirnov, {\it {Solar neutrinos and neutrino
  physics}},  {\em Eur. Phys. J.} {\bf A52} (2016), no.~4 87,
  [\href{http://arxiv.org/abs/1507.05287}{{\tt arXiv:1507.05287}}].

\bibitem{Hohenegger:2008zk}
A.~Hohenegger, A.~Kartavtsev, and M.~Lindner, {\it {Deriving Boltzmann
  Equations from Kadanoff-Baym Equations in Curved Space-Time}},  {\em Phys.
  Rev.} {\bf D78} (2008) 085027, [\href{http://arxiv.org/abs/0807.4551}{{\tt
  arXiv:0807.4551}}].

\bibitem{Garny:2009qn}
M.~Garny, A.~Hohenegger, A.~Kartavtsev, and M.~Lindner, {\it {Systematic
  approach to leptogenesis in nonequilibrium QFT: Self-energy contribution to
  the CP-violating parameter}},  {\em Phys. Rev.} {\bf D81} (2010) 085027,
  [\href{http://arxiv.org/abs/0911.4122}{{\tt arXiv:0911.4122}}].

\end{thebibliography}

\providecommand{\href}[2]{#2}\begingroup\raggedright
\endgroup

\end{document}